\documentclass[prd,aps,a4,
twocolumn,superscriptaddress,preprintnumbers,nofootinbib]{revtex4-1}
\usepackage{pslatex}
\usepackage[pdftex]{graphicx}
\usepackage{psfrag}
\usepackage{epsfig}
\usepackage{color}
\usepackage{cancel}
\usepackage{slashed}
\usepackage{amssymb}
\usepackage{amsmath}
\usepackage{hyperref}
\usepackage{enumerate}
\usepackage{multirow}
\usepackage{natbib}
\begin{document}
\title{Impact of form factor uncertainties on interpretations of \\ coherent 
elastic  neutrino-nucleus scattering data}
\author{D. Aristizabal Sierra}%
\email{daristizabal@ulg.ac.be}%
\affiliation{Universidad T\'ecnica
  Federico Santa Mar\'{i}a - Departamento de F\'{i}sica\\
  Casilla 110-V, Avda. Espa\~na 1680, Valpara\'{i}so, Chile}%
\affiliation{IFPA, Dep. AGO, Universit\'e de Li\`ege, Bat B5, Sart
  Tilman B-4000 Li\`ege 1, Belgium}%
\author{Jiajun Liao}%
\email{liaojiajun@mail.sysu.edu.cn}%
\affiliation{School of Physics, Sun Yat-Sen University, Guangzhou
  510275, China}%
\affiliation{Department of Physics and Astronomy, University of Hawaii
  at Manoa, Honolulu, HI 96822, USA}%
\author{D. Marfatia}%
\email{dmarf8@hawaii.edu}%
\affiliation{Department of Physics and Astronomy, University of Hawaii
  at Manoa, Honolulu, HI 96822, USA}%

\begin{abstract}
  The standard model coherent elastic neutrino-nucleus scattering
  (CE$\nu$NS) cross section is subject to nuclear form factor
  uncertainties, mainly driven by the root-mean-square radius of the
  neutron density distribution. Motivated by COHERENT phases I-III and
  future multi-ton direct detection dark matter searches, we evaluate
  these uncertainties in cesium iodide, germanium, xenon and argon
  detectors.  We find that the uncertainties become relevant for
  momentum transfers $q\gtrsim 20$~MeV and are essentially independent of
  the form factor parameterization. Consequently, form factor
  uncertainties are not important for CE$\nu$NS induced by reactor or
  solar neutrinos.  Taking into account these uncertainties, we then
  evaluate their impact on measurements of CE$\nu$NS at COHERENT, the
  diffuse supernova background (DSNB) neutrinos and sub-GeV
  atmospheric neutrinos. We also calculate the relative uncertainties in the 
  number of COHERENT events for different nuclei as a function of recoil energy. For
  DSNB and atmospheric neutrinos, event rates at a liquid argon detector  can be uncertain
  to more than 5\%. Finally, we consider the impact of form factor uncertainties on searches for
  nonstandard neutrino interactions, sterile neutrinos and neutrino
  generalized interactions. We point out that studies of new physics
  using CE$\nu$NS data are affected by neutron form factor
  uncertainties, which if not properly taken into account may lead to
  the misidentification of new physics signals. The uncertainties
  quantified here are also relevant for dark matter direct detection
  searches.
\end{abstract}

\maketitle
\section{Introduction}
\label{sec:intro}
Coherent elastic neutrino-nucleus scattering (CE$\nu$NS)
was observed by the COHERENT experiment in 2017 in a
cesium iodide scintillation detector. The measurement used neutrinos
produced at the spallation neutron source at the Oak Ridge National
Laboratory~\cite{Akimov:2017ade}. The cross section, obtained by the
coherent sum of the individual nucleon amplitudes, is the largest of
all neutrino cross sections at energies
$E_\nu\lesssim 100$~MeV, exceeding the elastic neutrino-electron
scattering cross section by about two orders of magnitude in typical
nuclei. The observation, however, relies on the detection of very small recoil
energies, which only recently became
possible with the use of the technology employed in direct detection
dark matter (DM) searches.

CE$\nu$NS data allow precise measurements of
the weak mixing angle~\cite{Canas:2018rng}, detailed studies of
nuclear structure through weak neutral current interactions~\cite{Patton:2012jr} and opens the possibility of searching for new
physics beyond the standard model (SM)~\cite{Barranco:2005yy,Scholberg:2005qs}. Indeed, since its
observation various studies of nonstandard neutrino
interactions (NSI)~\cite{Coloma:2017ncl,Liao:2017uzy,Kosmas:2017tsq,Billard:2018jnl},
sterile neutrinos~\cite{Billard:2018jnl}, neutrino generalized
interactions (NGI)~\cite{AristizabalSierra:2018eqm} and neutrino
electromagnetic properties~\cite{Kosmas:2017tsq,Cadeddu:2018dux} have
been presented. A proper interpretation of a CE$\nu$NS signal, as
related to any of these new physics scenarios, requires a detailed
understanding not only of experimental systematics errors but also of
theoretical uncertainties.

The calculation of event rates in CE$\nu$NS experiments involves proton and
neutron nuclear form factors, which account for the proton and neutron
distributions within the nucleus. In most treatments, however, the form factors
are assumed to be equal and so the nuclear form factor becomes a
global factor, which is typically parameterized in terms of the Helm~\cite{Helm:1956zz} form factor, the Fourier transform of the
symmeterized Fermi distribution~\cite{0305-4470-30-18-026}, or the
Klein-Nystrand form factor~\cite{Klein:1999qj} as adopted by the
COHERENT collaboration~ \cite{Akimov:2017ade}. These form factors
depend on various parameters whose values are fixed via experimental
data and so involve experimental uncertainties. Of particular
relevance is the root-mean-square (rms) radius of the nucleon
distribution, which in such analyses is fixed by, for example, the value
derived using a particular nuclear physics model or through a value
derived from fits to nuclear data~\cite{Lewin:1995rx}. This
simplification introduces an uncertainty on the predicted CE$\nu$NS
recoil spectrum (and number of events) in the SM as
well as in beyond-the-standard-model (BSM) physics scenarios.

The root-mean-square (rms) radius of the proton distribution
$r_\text{rms}^p$ is known from elastic electron-nucleus scattering
with a precision of order one-per-mille for nuclear isotopes up to
$Z=96$~\cite{Angeli:2013epw}. This is in sharp contrast with the rms
radius of the neutron distribution $r_\text{rms}^n$, which for almost
all nuclear isotopes is poorly known. Theoretical uncertainties on the
CE$\nu$NS process are therefore driven by the uncertainties in
$r_\text{rms}^n$. For the COHERENT experiment, the quenching factor
and neutrino flux uncertainties are of order 27\%~\cite{Akimov:2017ade,Akimov:2018vzs}. Thus,
form factor uncertainties are not particularly relevant in the
interpretation of current data. This situation, however, is expected
to change in the near future, and so form factor uncertainties will
play an important role.

Identifying the size of these uncertainties is crucial for
two reasons: (i) To understand whether a given signal is the result of new physics or of
 an ``unexpected'' nuclear physics effect, (ii)
DM direct detection in hundred-ton scale detectors like
Argo~\cite{Aalseth:2017fik}, will be subject to
irreducible neutrino backgrounds from the diffuse supernova background
(DSNB) and sub-GeV atmospheric neutrino fluxes. A precise
understanding of this background is crucial to discriminate between
neutrino-induced and WIMP-induced signals.

With this in mind, in this paper we investigate the size and behavior
of the neutron nuclear form factor uncertainties and their impact on
the interpretation of data. To that end we consider four
well-motivated nuclei: cesium iodide, germanium, xenon and
argon. The first three are (or will be) used by COHERENT in one of its
three phases~\cite{Akimov:2015nza}, while argon will be used by the
Argo detector of the Global Argon Dark Matter Collaboration which will
take 1000~ton-year of data~\cite{Aalseth:2017fik}. For definitiveness
we consider three nuclear form factor parameterizations: The Helm form
factor~\cite{Helm:1956zz}, the Fourier transform of the symmeterized
Fermi distribution~\cite{0305-4470-30-18-026} and the Klein-Nystrand
form factor~\cite{Klein:1999qj}. And we assume the same
parameterization for both protons and neutrons. We first study the size
and momentum transfer ($q$) dependence of the neutron form factor
uncertainties using these three parameterizations. After precisely
quantifying them, we study their impact in COHERENT and in an
argon-based multi-ton DM detector. We assess as well the impact of the
uncertainties on the interpretation of new physics effects. We do this
in the case of NSI, active-sterile neutrino oscillations in the 3+1
framework and spin-independent NGI. We evaluate the effects of the
neutron form factor uncertainties on the available parameter space and
the potential misidentification of new physics signals when these
uncertainties are not properly accounted for.

The paper is organized as follows. In Section~\ref{sec:cevns} we
introduce our notation and briefly discuss the CE$\nu$NS process,
focusing on form factor parameterizations and the corresponding rms
radii of the nucleon density distributions. In
Section~\ref{sec:ff-uncertainties} we quantify the size of the neutron
 form factor uncertainties, study their $q$ dependence
and show that they are fairly independent of the choice of the nuclear form factor. 
In Sections~\ref{sec:implications-1} and~\ref{sec:new-physics} 
we study the implications for
SM predictions and for new physics
searches, respectively. In Section~\ref{sec:conclusions} we present our conclusions. In
Appendix~\ref{sec:DSNB} we present the details of the calculation
of the DSNB neutrino flux, while in Appendix~\ref{sec:parameters-def} we provide
details of the NGI analysis.
\section{Coherent elastic neutrino-nucleus scattering}
\label{sec:cevns}
For neutrino energies below $\sim 100$\,MeV the de Broglie wavelength
of the neutrino-nucleus process is larger than the typical nuclear
radius and so the individual nucleon amplitudes add coherently. In the
SM this translates into a cross section that is approximately enhanced by
the number of constituent neutrons $N$~\cite{Freedman:1973yd,Freedman:1977xn}:
\begin{equation}
  \label{eq:cevns-xsec}
  \frac{d\sigma}{dE_r}=\frac{G_F^2m_N}{2\pi}
  \left(2-\frac{E_rm_N}{E_\nu^2}\right)
  \left[Ng_V^nF_N(q^2)+Zg_V^pF_Z(q^2)\right]^2\,,
\end{equation}
where $E_r=q^2/2m_N$ is the nuclear recoil energy.
This result follows from the vector neutral current. 
The axial current contribution, being
spin dependent, is much smaller. The neutron and proton charges are
given by $g_V^n=-1/2$ and $g_V^p=1/2-2\sin^2\theta_W$, with $\theta_W$
the weak mixing angle. In the Born approximation, the nuclear
form factors $F_{N,Z}(q^2)$ follow from the Fourier transform of the
neutron and proton density distributions. They capture the behavior one
expects: the cross section should fall with increasing neutrino energy
(increasing $q$). Theoretical predictions based on
Eq.~(\ref{eq:cevns-xsec}) involve uncertainties from electroweak
parameters and nuclear form factors. These
uncertainties should be accounted for and are particularly important
in searches for new physics effects, which arguably are not expected
to significantly exceed the SM expectation. Since the uncertainty in $G_F$ is
a few tenths of a part per million~\cite{Tishchenko:2012ie},
electroweak uncertainties are dominated by the weak mixing angle for
which (using the $\overline{\text{MS}}$ renormalization scheme at
the $Z$ boson mass scale) the PDG gives~\cite{Tanabashi:2018oca}
\begin{equation}
  \label{eq:EW-uncertainties}
  \sin^2\theta_W=0.23122\pm 0.00003 \,.
\end{equation}
Electroweak uncertainties are therefore of no relevance. On the
contrary, since nuclear form factors encode information on the proton
and neutron distributions one expects these uncertainties to be
sizable and more pronounced for large $q$, given the behavior
$F(q^2\to 0)\to 1$.
These uncertainties turn out to be crucial for the
interpretation of data from fixed target experiments such as
COHERENT~\cite{Akimov:2017ade,Akimov:2018vzs} and for DM direct
detection experiments subject to diffuse supernova background (DSNB)
and sub-GeV atmospheric neutrinos~\cite{Billard:2013qya}.
\subsection{Nuclear form factors}
\label{sec:nuclear-FFs}
Form factors are introduced to account for the
density distributions of nucleons inside the nucleus. They follow from the Fourier
transform of the nucleon distributions,
\begin{equation}
  \label{eq:fourier-transform}
  F(q^2)=\int\,e^{i\vec{q}\cdot \vec{r}}\,\rho(r)\,d^3\vec{r}\ .
\end{equation}
The basic properties of nucleonic distributions are captured by
different \textit{parameterizations}. Here we consider those provided
by the Helm model~\cite{Helm:1956zz}, the symmeterized Fermi
distribution \cite{0305-4470-30-18-026} and the Klein-Nystrand
approach \cite{Klein:1999qj}. These distributions depend on two
parameters which measure different nuclear properties and which are
constrained by means of the rms radius of the distribution,
\begin{equation}
  \label{eq:rrms}
  r_\text{rms}^2\equiv\langle r^2\rangle=
  \frac{\int\,\rho(r)\,r^2\,d^3\vec{r}}
  {\int\,\rho(r)\,d^3\vec{r}}\ .
\end{equation}
In what follows we briefly discuss these parameterizations and the
relations between their defining parameters and the rms radius of the
distributions. These relations are key to our analysis for they
determine, through  the  experimental uncertainties in $r_\text{rms}$, the
extent up to which these parameters can vary, thereby defining the form
factor uncertainties.

In the Helm model the nucleonic distribution is given by a convolution
of a uniform density with radius $R_0$ (box or diffraction radius) and
a Gaussian profile. The latter is characterized by the folding width
$s$, which accounts for the surface thickness. Thus, the Helm
distribution reads
\begin{equation}
  \label{eq:Helm-dist}
  \rho_H(r)={3 \over 4\pi R_0^3}\,\int\,f_G(r - r')\theta(R_0-|r'|)
  d^3\vec{r'}\ ,
\end{equation}
with $\theta(x)$ a Heaviside step function and $f_G(x)$ a Gaussian
distribution given by
\begin{equation}
  \label{eq:gaussian-helm}
  f_G(x)=\frac{e^{-x^2/2s^2}}{\sqrt{(2\pi)^3}s^3}\,.
\end{equation}
 The Helm form factor is then derived from
Eqs.~(\ref{eq:fourier-transform}) and~(\ref{eq:Helm-dist}):
\begin{equation}
  \label{eq:Helm}
  F_\text{H}(q^2)=3\frac{j_1(qR_0)}{qR_0}e^{-q^2s^2/2}\,,
\end{equation}
where $j_1(x)=\sin(x)/x^2-\cos(x)/x$ is a spherical Bessel function of
order one. The rms radius is obtained from Eqs.~(\ref{eq:rrms}) and (\ref{eq:Helm-dist}) and is
given by
\begin{equation}
  \label{eq:first-momentum-Helm}
  \langle r^2 \rangle_\text{H}=\frac{3}{5}R_0^2+3s^2\,.
\end{equation}

The symmeterized Fermi density distribution $\rho_\text{SF}(r)$
follows from the symmeterized Fermi function
$f_\text{SF}(r)=f_\text{F}(r)+f_\text{F}(-r)-1$,
which in turn follows from the conventional Fermi, or Woods-Saxon
function,
\begin{equation}
  \label{eq:fermi-distribution}
  f_\text{F}(r)=\frac{1}{1+e^{(r-c)/a}}\ ,
\end{equation}
where $c$ is the half-density radius and $a$ represents the
surface diffuseness. Accordingly, $\rho_\text{SF}(r)$ can be
written as
\begin{align}
  \label{eq:sff-rho}
  \rho_\text{SF}(r)
  &=\frac{3}{4\pi c(c^2+\pi^2 a^2)}
  \frac{\sinh(c/a)}{\cosh(r/a)+\cosh(c/a)}\,.
\end{align}
 In contrast to the Fermi
density distribution it has the advantage that its Fourier transform
can be analytically evaluated with the result,
\begin{align}
  \label{eq:SF}
  F_\text{SF}(q^2)&=\frac{3}{qc}\left[\frac{\sin(qc)}{(qc)^2}
  \left(\frac{\pi q a}{\tanh(\pi q a)}\right)-\frac{\cos(qc)}{qc}\right]      
  \nonumber\\
  &\times \left(\frac{\pi q a}{\sinh(\pi q a)}\right) \frac{1}{1+(\pi a/c)^2}\ .
\end{align}
Then,
\begin{equation}
  \label{eq:rrms-sff}
  \langle r^2 \rangle_\text{SF}
  =\frac{3}{5}c^2+\frac{7}{5}(\pi a)^2\ .
\end{equation}

The Klein-Nystrand approach relies on a surface-diffuse distribution
which results from folding a short-range Yukawa potential with range
$a_k$, over a hard sphere distribution with radius $R_A$. The Yukawa
potential and the hard sphere distribution can be written as
\begin{equation}
  \label{eq:Yukawa-pot}
  V_\text{Y}(r)=\frac{e^{-r/a_k}}{4\pi a_k^2r}\ ,\qquad
  \rho_\text{HS}(r)=\frac{3}{4\pi R_A^3}\theta(R_A-r)\ .
\end{equation}
The Klein-Nystrand form factor can then be calculated as the product
of two individual Fourier transformations, one of the potential and
another of the hard sphere distribution, resulting in
\begin{equation}
  \label{eq:KN-FF}
  F_\text{KN}(q^2)=F_\text{Y}(q^2)F_\text{HS}(q^2)
  =3\frac{j_1(qR_A)}{qR_A}\frac{1}{1+q^2a_k^2}\ .
\end{equation}
The rms radius is given by
\begin{equation}
  \label{eq:rrms-KN}
    \langle r^2 \rangle_\text{KN}=\frac{3}{5}R_A^2+6a_k^2\ .
\end{equation}

\section{Form factor uncertainties}
\label{sec:ff-uncertainties}
The rms radii of the proton density distributions are determined from
different experimental sources. The values reported in
\cite{Angeli:2013epw} include data from optical and $K_\alpha$ X-ray
isotope shifts as well as muonic spectra and electronic scattering
experiments. This wealth of data has allowed the determination of
$\sqrt{\langle r^2_p \rangle}\equiv r_\text{rms}^p$ with high accuracy
for all isotopes of interest for CE$\nu$NS and DM direct detection
experiments. The rms radii for the proton distribution are as in
Table~\ref{tab:nuclei-parameters}. In contrast, rms radii of the
neutron density distributions
$\sqrt{\langle r^2_n\rangle}\equiv r_\text{rms}^n$ are poorly known,
mainly because barring the cases of $^{208}$Pb, $^{133}$Cs and
$^{127}$I~\cite{Abrahamyan:2012gp,Horowitz:2012tj,Horowitz:2013wha,Cadeddu:2017etk},
their experimental values follow from hadronic experiments which are
subject to large uncertainties.

\begin{table*}
  \centering
  \renewcommand{\arraystretch}{1.6}
  \begin{tabular}{|c|c|c||cc|c||cc|c|cc|c||cc|c|cc|c|cc|c|}\hline
    &\multicolumn{2}{|c||}{}
    &\multicolumn{3}{|c||}{Argon}
    &\multicolumn{6}{|c||}{Germanium}
    &\multicolumn{9}{|c|}{Xenon}\\\hline\hline
    \multirow{3}{*}
    &{\footnotesize$^{127}$I}&{\footnotesize$4.750$}
    &{\footnotesize$^{36}$Ar}&{\footnotesize($0.33\%$)}&{\footnotesize$3.390$}
    &{\footnotesize$^{70}$Ge}&{\footnotesize($20.4\%$)}&{\footnotesize$4.041$}
    &{\footnotesize$^{72}$Ge}&{\footnotesize($27.3\%$)}&{\footnotesize$4.057$}
    &{\footnotesize$^{124}$Xe}&{\footnotesize($0.095\%$)}&{\footnotesize$4.766$}
    &{\footnotesize$^{126}$Xe}&{\footnotesize($0.089\%$)}&{\footnotesize$4.774$}
    &{\footnotesize$^{128}$Xe}&{\footnotesize($1.91\%$)}&{\footnotesize$4.777$}\\\cline{2-21}
    &{\footnotesize$^{133}$Cs}&{\footnotesize$4.804$}
    &{\footnotesize$^{38}$Ar}&{\footnotesize($0.06\%$)}&{\footnotesize$3.402$}
    &{\footnotesize$^{73}$Ge}&{\footnotesize($7.76\%$)}&{\footnotesize$4.063$}
    &{\footnotesize$^{74}$Ge}&{\footnotesize($36.7\%$)}&{\footnotesize$4.074$}
    &{\footnotesize$^{129}$Xe}&{\footnotesize($26.4\%$)}&{\footnotesize$4.777$}
    &{\footnotesize$^{130}$Xe}&{\footnotesize($4.07\%$)}&{\footnotesize$4.781$}
    &{\footnotesize$^{131}$Xe}&{\footnotesize($21.2\%$)}&{\footnotesize$4.780$}\\\cline{2-21}
    &{\footnotesize---}&{\footnotesize---}
    &{\footnotesize$^{40}$Ar}&{\footnotesize($99.6\%$)}&{\footnotesize$3.427$}
    &{\footnotesize$^{76}$Ge}&{\footnotesize($7.83\%$)}&{\footnotesize$4.09$}
    &{\footnotesize---}&{\footnotesize---}&{\footnotesize---}
    &{\footnotesize$^{132}$Xe}&{\footnotesize($26.9\%$)}&{\footnotesize$4.785$}
    &{\footnotesize$^{134}$Xe}&{\footnotesize($10.4\%$)}&{\footnotesize$4.789$}
    &{\footnotesize$^{136}$Xe}&{\footnotesize($8.86\%$)}&{\footnotesize$4.796$}\\\cline{1-21}
  \end{tabular}
  \caption{Rms radii (in fm) of the proton density distributions of the stable isotopes of cesium, 
    iodine, argon, germanium and xenon~\cite{Angeli:2013epw}. The relative abundances of the Ar, Ge and Xe isotopes are
    provided in parentheses.}
  \label{tab:nuclei-parameters}
\end{table*}
At the form factor level, therefore, uncertainties on $r_\text{rms}^p$
are basically irrelevant while uncertainties in $r_\text{rms}^n$ have
a substantial effect. Consequently, we adopt the following
procedure. We verified that adopting different form factor
parameterizations for the proton and neutron distributions leads to a
small effect on our results, so we assume the same form factor for
both. For protons, in each of Eqs.~(\ref{eq:first-momentum-Helm}),
(\ref{eq:rrms-sff}) and (\ref{eq:rrms-KN}), we fix one parameter and
determine the other by fixing $r_\text{rms}^p$ to its experimental
central value. For neutrons we do the same as that for protons, but
restrict $r_\text{rms}^n$ to values above $r_\text{rms}^p$; this lower
limit is reliable provided $N>Z$, which is the case for all
  nuclei we consider. For nuclei other than argon, we fix the upper limit using the neutron
skin, $\Delta r_{np}= r_\text{rms}^n-r_\text{rms}^p$, of $^{208}$Pb,
which is measured by the PREX experiment at Jefferson laboratory to be
$\Delta
r_{np}(^{208}\text{Pb})=0.33^{+0.16}_{-0.18}$~fm~\cite{Horowitz:2012tj,Horowitz:2013wha}; while
PREX-II and CREX will measure the neutron skins of $^{208}\text{Pb}$
and $^{48}\text{Ca}$, respectively~\cite{Horowitz:2013wha}, no
measurements of the neutron skin of the nuclei we are considering
are planned. Experiments have focused on the doubly-magic nuclei,
$^{208}$Pb and $^{48}$Ca, because for such nuclei theoretical
calculations are under relatively good control. Pairing correlations
and deformation become relevant for nuclei that are not
doubly-magic. The situation is worse for nuclei with unpaired nucleons
like $^{133}$Cs, $^{127}$I and $^{129}$Xe, in which case calculations
assume that the nuclei are even-even nuclei (although they are not),
and rescale the occupancy of the valence orbital by a suitable factor
with the hope that bulk properties like the weak radius are not
sensitive to this ``spherical approximation''~\cite{Jorge}.

We then require the neutron skin of the heavy nuclei to be no larger
than 0.3~fm given that their values of $(N-Z)/(N+Z)$ are less than for
$^{208}\text{Pb}$.  We use
\begin{equation}
  r_{\rm{rms}}^n|_{\rm max}\equiv r_{\rm{rms}}^p+0.3~{\rm fm}\  \  \  \text{for Cs, I, Ge and Xe.}
\end{equation}
For argon, we allow the neutron skin to lie between 0.1~fm and 0.2~fm, i.e.,
\begin{equation}
  r_{\rm{rms}}^n|_{\rm max}\equiv r_{\rm{rms}}^p+0.1~\text{fm\ \ \  or\ \ \ }~r_{\rm{rms}}^p+0.2~{\rm fm}\  \  \  \text{for Ar.}
\end{equation}
Note that these large values of $\Delta r_{np}$ parameterize the
envelope of the form factors from different calculation methods including chiral effective field theory,
relativistic and  nonrelativistic mean-field models, etc.   It
is a proxy for the spread in theoretical predictions of the form
factor, and is not intended as an estimate of its value.

\begin{figure*}
  \centering
  \includegraphics[scale=0.35]{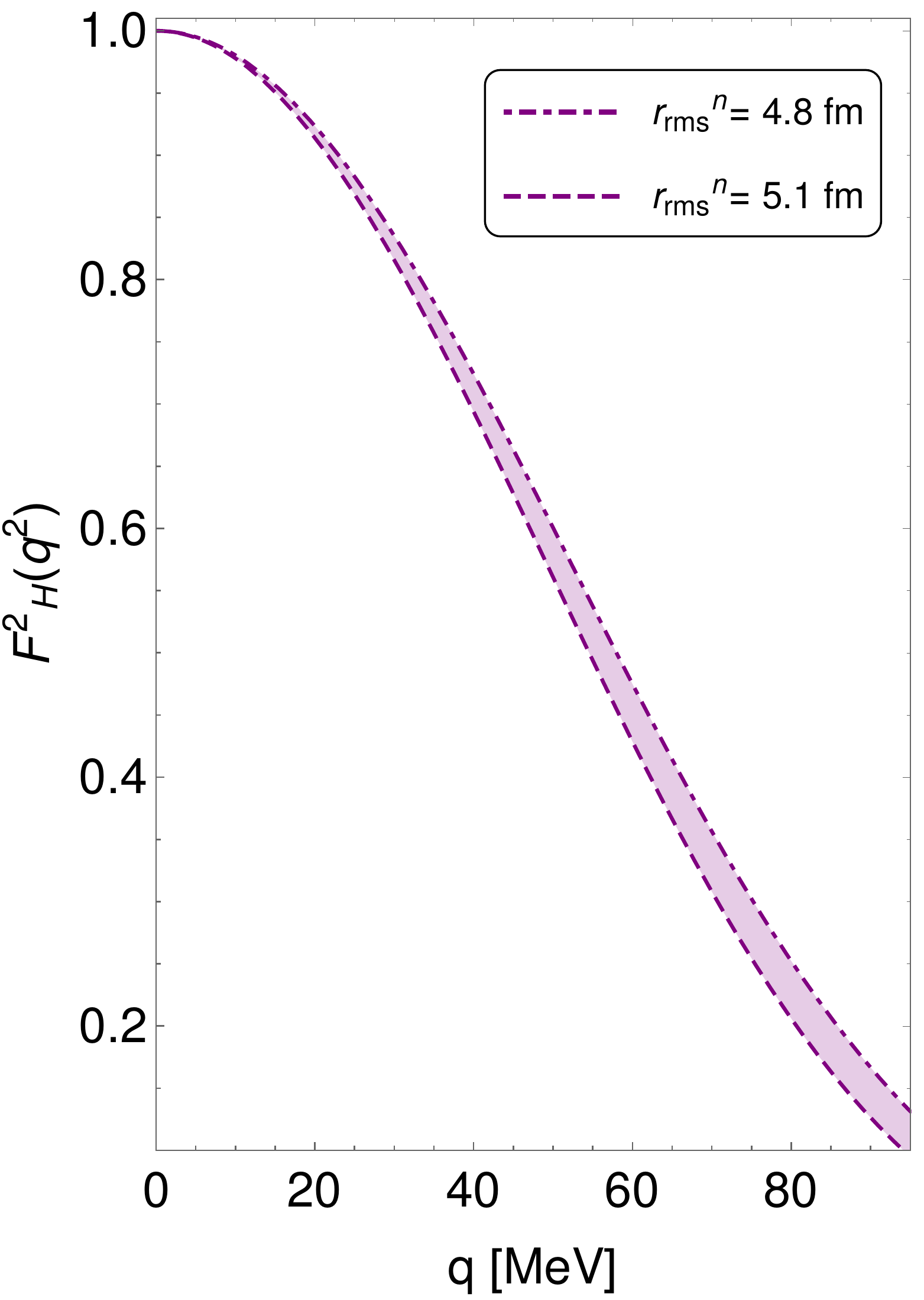}
  \hspace{0.5cm}
  \includegraphics[scale=0.333]{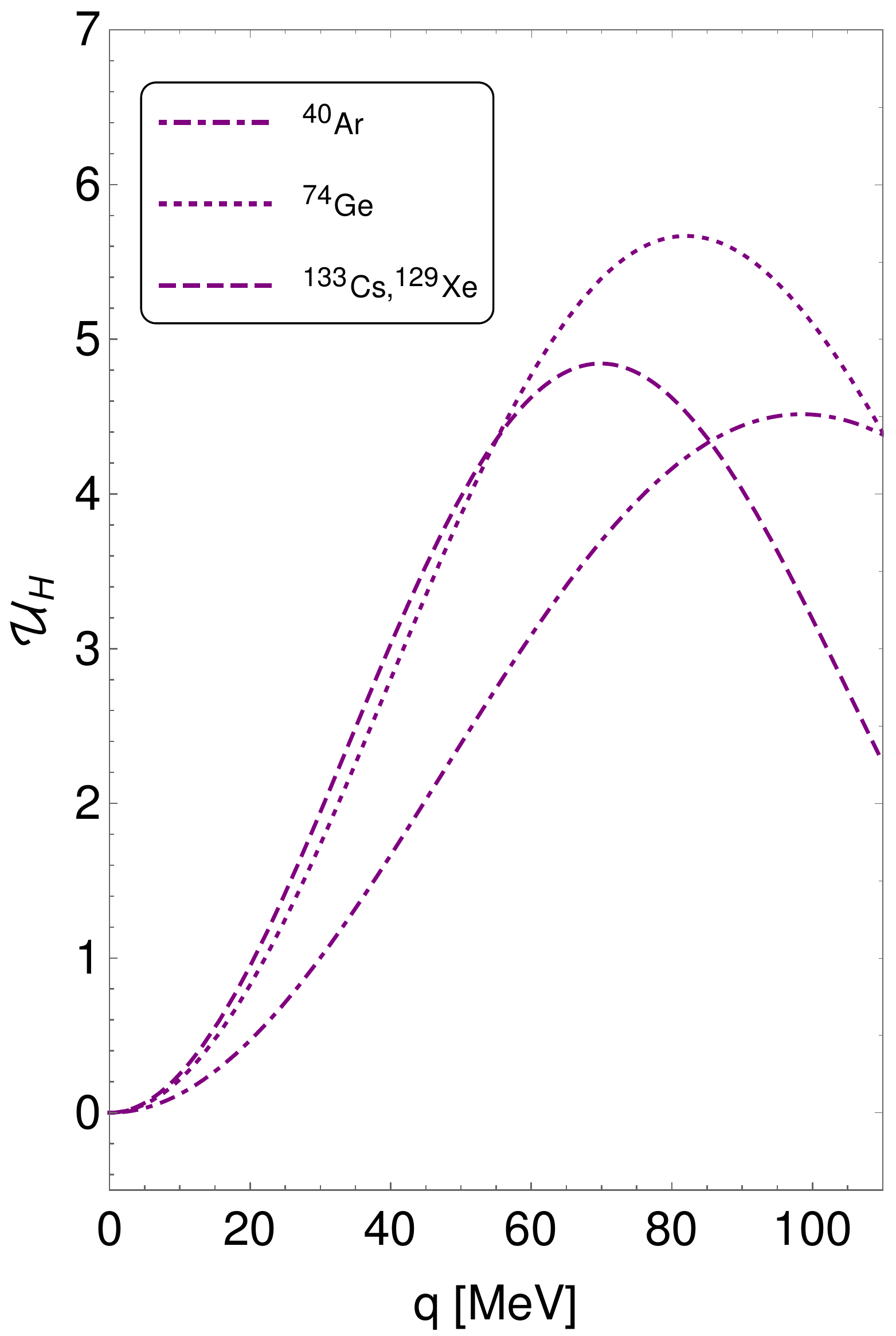}
  \hspace{0.5cm}
  \includegraphics[scale=0.35]{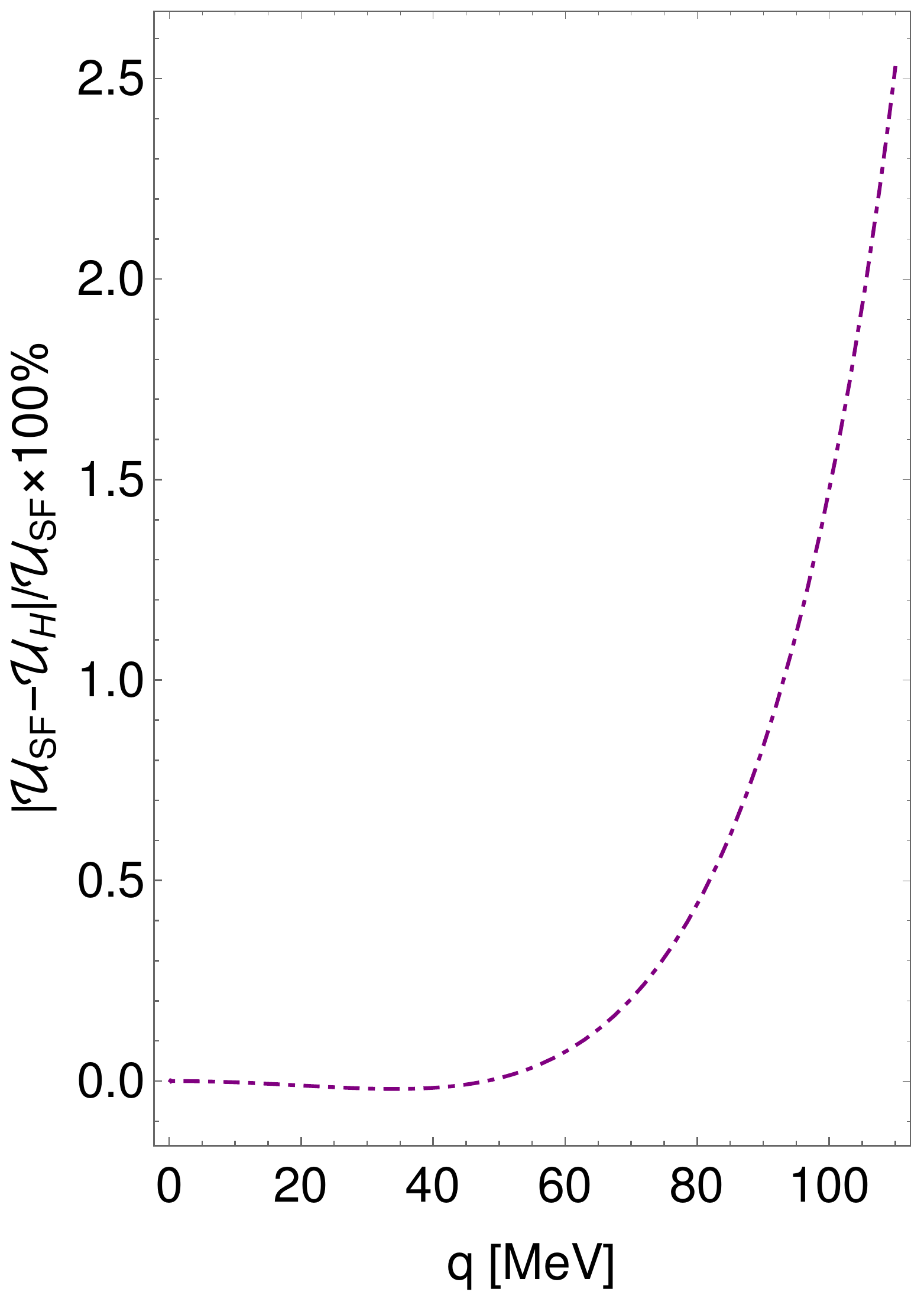}
  \caption{\textbf{Left}: Square of the Helm form factor as a function
    of the momentum transfer $q$. The upper (lower) curve is obtained
    by fixing $\sqrt{\langle r^2\rangle_\text{H}}$ to $r_\text{rms}^p$
    ($r_\text{rms}^p+0.3$~fm) as given in
    Table~\ref{tab:nuclei-parameters} for $^{133}$Cs. \textbf{Middle}:
    Percentage uncertainty for $^{133}$Cs, $^{129}$Xe, $^{74}$Ge and
    $^{40}$Ar (for $^{40}$Ar, we take
    $r_{\rm{rms}}^n|_{\rm max}=r_\text{rms}^p+0.2$~fm).  It can be seen that uncertainties get
    larger as $q$ increases to 65~MeV. \textbf{Right}: Relative
    difference between the Helm and symmeterized Fermi form factors
    uncertainties for $^{133}\text{Cs}$. Using the Klein-Nystrand form
    factor yields results of the same order and so are not
    displayed. The main point is that the size of the uncertainties do
    not depend on the form factor chosen.}
  \label{fig:Helm-vs-q-and-percentage-uncertainty}
\end{figure*}
For the Helm form factor we fix the surface thickness $s$ to 0.9
fm~\cite{Lewin:1995rx}, for the form factor based on the symmeterized
Fermi function we fix the surface diffuseness $a$ to 0.52
fm~\cite{Piekarewicz:2016vbn} and for the Klein-Nystrand form factor
we fix the range $a_k$ of the Yukawa potential to
0.7~fm~\cite{Klein:1999qj}. We checked that our results are rather
insensitive to variations of these values.  With the procedure already
outlined, we first investigate the behavior of the uncertainties and
their size. Figure~\ref{fig:Helm-vs-q-and-percentage-uncertainty}
shows the result for the Helm form factor obtained for $^{133}$Cs. The
left panel shows that for low $q$, the uncertainties are small and
increase with increasing momentum, reaching their maximum for
$q\simeq 65$~MeV. This behavior is apparent in the middle panel which
shows the Helm percentage uncertainty,
\begin{equation}
  \label{eq:percentage-uncertainty}
  \mathcal{U}_H=\left|F_H^2(q^2)|_{r_\text{rms}^n=r_\text{rms}^p}
    -F_H^2(q^2)|_{r_\text{rms}^n=r_\text{rms}^p+0.3\,\text{fm}}\right|\times 100\%\ ,
\end{equation}
which measures the size of the spread due to the uncertainties in
$r_\text{rms}^n$; for argon, 0.3~fm is replaced by 0.2~fm in the above equation. It can be seen that in the case of $^{133}$Cs the
uncertainty can be as large as 5\%, and for $^{40}$Ar as large as
$4.5\%$.  To address how this result depends on the choice of form
factor, we calculate the percentage uncertainty for $F_\text{SF}$ and
$F_\text{KN}$, with the aid of Eqs.~(\ref{eq:rrms-sff}) and
(\ref{eq:rrms-KN}). The right panel in
Fig.~\ref{fig:Helm-vs-q-and-percentage-uncertainty} shows the relative
uncertainty obtained by comparing the uncertainties from the Helm and
the symmeterized Fermi form factors, calculated according to
$|\mathcal{U}_\text{SF}-\mathcal{U}_\text{H}|/\mathcal{U}_\text{SF}\times
100\%$;
results using the Klein-Nystrand form factor are similar and are not
displayed.  It can be seen that uncertainties are parameterization
independent for $q$ up to 60 MeV or so. For larger $q$, differences
are at most of order $2.5\%$, with the Helm form factor yielding
slightly larger values. In summary, the conclusions derived from
Fig.~\ref{fig:Helm-vs-q-and-percentage-uncertainty} hold no matter the
form factor choice. Henceforth, to calculate the impact of the form
factor uncertainties on CE$\nu$NS, we employ the Helm form factor.
\section{Implications for COHERENT, DSNB and sub-GeV atmospheric
  neutrinos}
\label{sec:implications-1}
We now turn to the study of the impact of the form factor
uncertainties on SM predictions for CE$\nu$NS. We
begin with COHERENT in each of its phases. For phase-I we
calculate the expected number of events taking into account the
contributions from both $^{133}$Cs and $^{127}$I. For phase-II
(germanium phase) and phase-III (LXe phase) we calculate the number of
events assuming the specifications given in \cite{Akimov:2015nza} with
the number of protons on target
($n_\text{POT}$) per year as in the CsI case; event numbers for a different value
$n_\text{POT}^\prime$ can be straightforwardly rederived by scaling
our result by $n_\text{POT}^\prime/n_\text{POT}$. Since germanium and
xenon have several sufficiently abundant isotopes (see
Table~\ref{tab:nuclei-parameters}), we calculate the recoil spectrum generated by each
of the nuclides.  The $i^{\rm{th}}$ isotope recoil spectrum can be written as
\begin{equation}
  \label{eq:recoil-spectrum}
  \frac{dR_i}{dE_r}=\frac{m_\text{det}N_A}{\langle m\rangle}X_i
  \int_{E_\nu^\text{min}}^{E_\nu^\text{max}}\,\phi(E_\nu)
  \,\frac{d\sigma_i}{dE_r}\,dE_\nu\ ,
\end{equation}
where $m_\text{det}$ refers to the detector mass, $X_i$ to
its relative abundance, $\langle m\rangle$ to the average molar mass
calculated as $\sum_k m_k X_k$ ($m_k$ being the molar mass of the
individual isotopes), $N_A=6.022\times 10^{23}\,\text{mol}^{-1}$ and
$\phi(E_\nu)$ the neutrino flux. Note
that the global factor $(m_\text{det}N_A/\langle m\rangle) X_i$
corresponds to the number of nuclei of the $i^{\rm{th}}$ type in the
detector.  The differential cross section is given by
Eq.~(\ref{eq:cevns-xsec}) with $m_N\to m_i$, $N\to N_i$ and
$q^2\to q^2_i=2m_i E_r$. For each isotope contribution
$r_\text{rms}^p$ is fixed according to the values in
Table~\ref{tab:nuclei-parameters} and $r_\text{rms}^n$ as
described in the previous section. Calculating the individual
recoil spectra according to Eq.~(\ref{eq:recoil-spectrum}) and then
summing over all of them (to determine the total recoil spectrum),
allows to properly trace the uncertainties induced by each neutron
form factor.

For the COHERENT phase-I analysis we use
$m_\text{det}=14.6$~kg and adapt
Eq.~(\ref{eq:recoil-spectrum}) to take into account the contributions
of $^{133}$Cs and $^{127}$I. This is done by trading $X_i$ for the
nuclear fractions $f_i=A_i/(A_\text{Cs}+A_\text{I})$ ($A_i$ refer to
the $^{133}$Cs and $^{127}$I mass numbers) in
(\ref{eq:recoil-spectrum}) and $\langle m \rangle$ for
$m_\text{CsI}=259\times 10^{-3}\,$kg/mol (CsI molar mass). 
The acceptance function is~\cite{Akimov:2018vzs}
\begin{equation}
  \label{eq:acceptance}
  \mathcal{A}(n_\text{PE})=\frac{k_1}{1+e^{-k_2(n_\text{PE}-x_0)}}\theta(n_\text{PE}-5)\ ,
\end{equation}
where $k_1=0.6655$, $k_2=0.4942$, $x_0=10.8507$, and $n_\text{PE}$ is
the observed number of photoelectrons.\footnote{For
    the CsI COHERENT analysis we use the relation
    $n_\text{PE}=1.17(E_r/\text{keV})$. For the
    germanium, xenon and argon detectors we use
    Heaviside step functions with 2~keV, 5~keV and 20~keV thresholds,
    respectively, and display the results as a function
    of recoil energy.} Neutrino fluxes in COHERENT are produced by
$\pi^+$ and $\mu^+$ decays, and so three neutrino flavors are produced
($\nu_\mu$, $\bar\nu_\mu$ and $\nu_e$) with known energy spectra:
\begin{align}
  \label{eq:nu-spectra.COHERENT}
  \mathcal{F}_{\nu_\mu}(E_{\nu_\mu})&=
  \frac{2m_\pi}{m_\pi^2-m_\mu^2}\,
  \delta\left(
  1-\frac{2E_{\nu_\mu}m_\pi}{m_\pi^2-m_\mu^2}
  \right) \ ,
  \nonumber\\
  \mathcal{F}_{\nu_e}(E_{\nu_e})&=\frac{192}{m_\mu}
  \left(\frac{E_{\nu_e}}{m_\mu}\right)^2
  \left(\frac{1}{2}-\frac{E_{\nu_e}}{m_\mu}\right)\ ,
  \nonumber\\
  \mathcal{F}_{\bar\nu_\mu}(E_{\bar\nu_\mu})&=\frac{64}{m_\mu}
  \left(\frac{E_{\bar\nu_\mu}}{m_\mu}\right)^2
  \left(\frac{3}{4}-\frac{E_{\bar\nu_\mu}}{m_\mu}\right)\,,
\end{align}
where the neutrino energies are less than $m_\mu/2$.

\begin{figure*}
  \centering
  \includegraphics[scale=0.4]{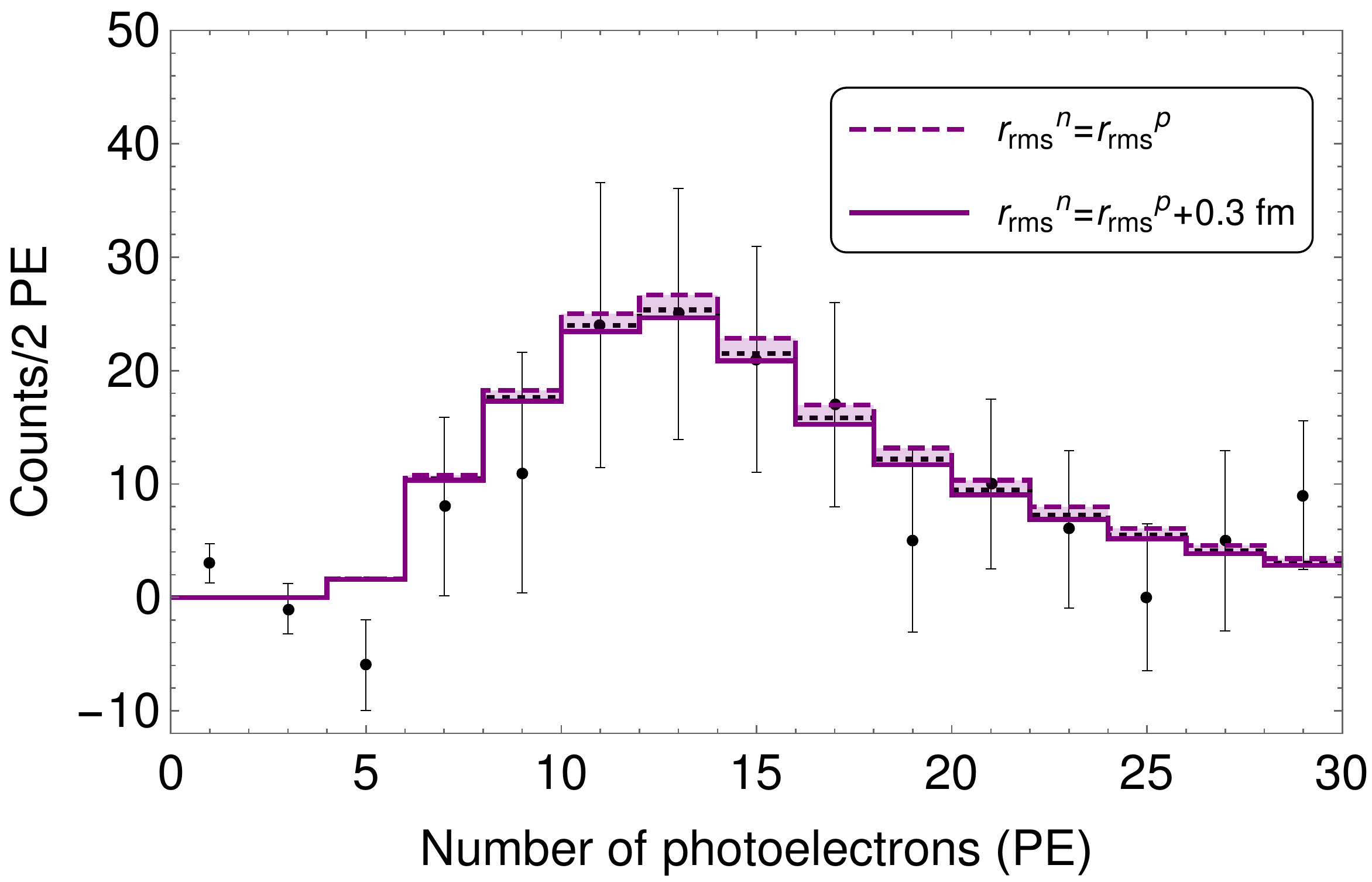}
  \caption{The expected CE$\nu$NS residual event
    spectrum in COHERENT as a function of the number of photoelectrons
    as predicted in the SM. The points correspond to COHERENT data
    (with their error bars) \cite{Akimov:2017ade}, while the shaded
    region between the two histograms defines the uncertainty in the
    spectrum due to the form factor uncertainties. The black-dotted
    histogram is obtained by fixing the rms radii of the $^{133}$Cs
    and $^{127}$I neutron density distributions to values obtained
    from theoretical calculations; see
    Eq.~(\ref{eq:rnrms-nuclear-physics-prediction}). For the dashed
    histogram, $r_\text{rms}^n=r_\text{rms}^p$, with the values of
    $r_\text{rms}^p$ from Table~\ref{tab:nuclei-parameters}. The Helm
    form factor has been used but a different form factor does not
    noticeably alter the spectra.}
\label{fig:coherent}
\end{figure*}
The neutrino flux per flavor $\phi_a(E_\nu)$ at the detector is
obtained by weighting the energy spectra by the normalization factor
$\mathcal{N}=r\times n_\text{POT}/4\pi L^2$. Here $r=0.08$ determines
the number of neutrinos produced per proton collision (per flavor),
$L=19.3$~m is the distance from the source to the detector, and
$n_\text{POT}=1.76\times 10^{23}$ is the number of protons on target
in the 308.1 days of neutrino production~\cite{Akimov:2017ade}, which
corresponds to $2.1\times 10^{23}\,\text{protons}/\text{year}$. In
terms of $n_\text{PE}$ we calculate the SM expectation for the number
of events per bin (2 photoelectrons) taking into account the neutron
form factor uncertainties. The result is displayed in
Fig.~\ref{fig:coherent}. Uncertainties in the neutron form factor
produce an uncertainty in the expected number of events, with a
behavior such that small values of $r_\text{rms}^n$ tend to increase
the number of events, while large values tend to decrease them.  This
is in agreement with the result in the left panel in
Fig.~\ref{fig:Helm-vs-q-and-percentage-uncertainty}. One can see as
well that for low $n_\text{PE}$ (recoil energy), no sizable
uncertainties are observed. However, for $n_\text{PE}=7$
($E_r=5.98$~keV) uncertainties are of order
4\% and increase to about 9\% for
$n_\text{PE}=15$.  We also calculate the number of events by fixing
the rms radii of the $^{133}$Cs and $^{127}$I neutron density
distributions to
\begin{equation}
  \label{eq:rnrms-nuclear-physics-prediction}
  r_\text{rms}^n(^{133}\text{Cs})=5.01\ \text{fm}\,,
  \qquad
  r_\text{rms}^n(^{127}\text{I})=4.94\ \text{fm}\,.
\end{equation}
These values follow from theoretical calculations using the
relativistic mean field (RMF) NLZ2 nuclear
model~\cite{Cadeddu:2017etk}. The result obtained can then be regarded
as purely theoretical. Comparing the black-dotted histogram in
Fig.~\ref{fig:coherent} with those determined by the form factor
uncertainties we see that the theoretical expectation is closer to the
result for $r_\text{rms}^p$. We have checked that because of the large
experimental uncertainties, using different values of $r_\text{rms}^n$
has almost no effect on the quality of the fit.

\begin{figure*}
  \centering
   \includegraphics[scale=0.37]{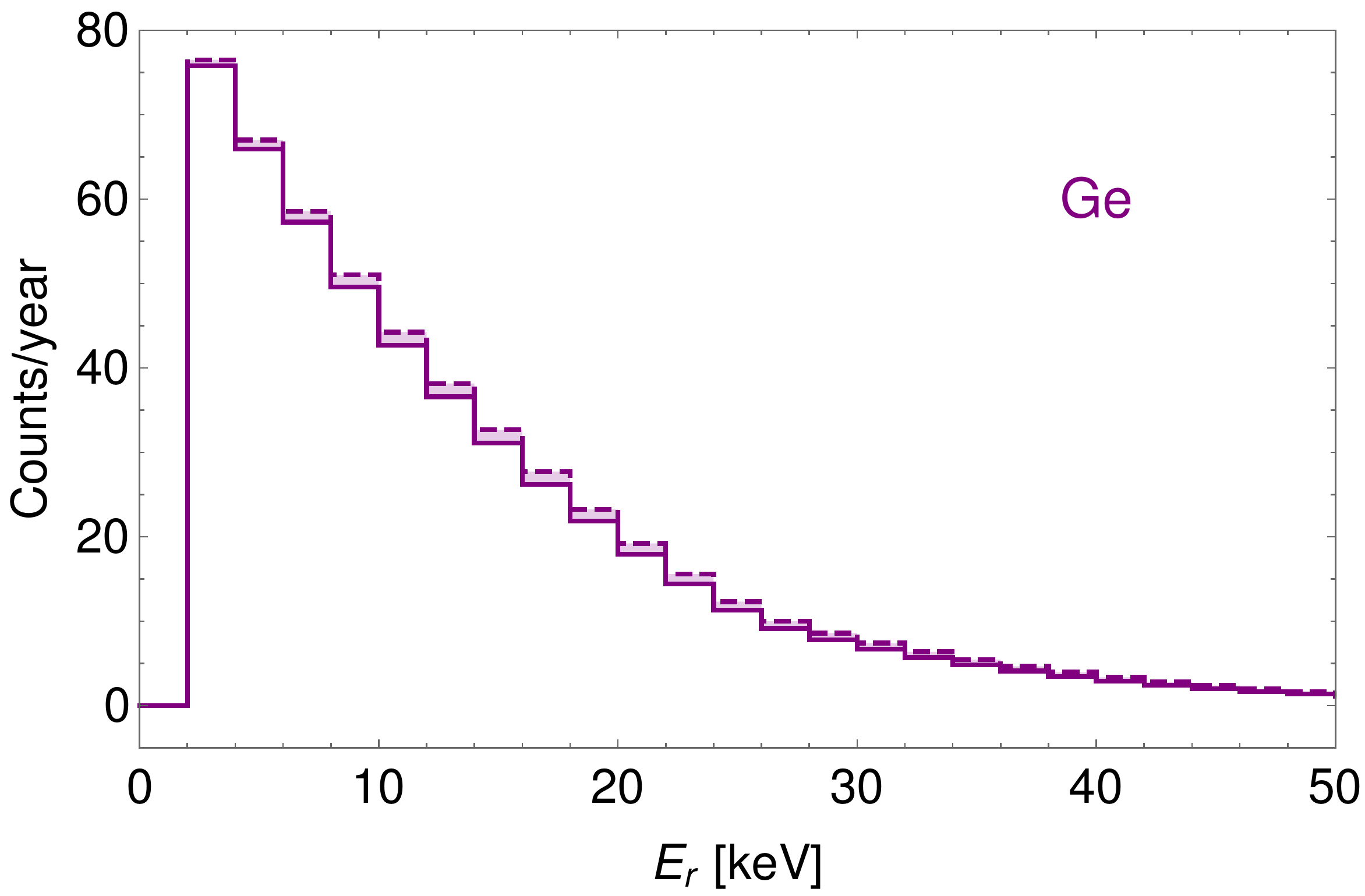}
  \hfill
  \includegraphics[scale=0.365]{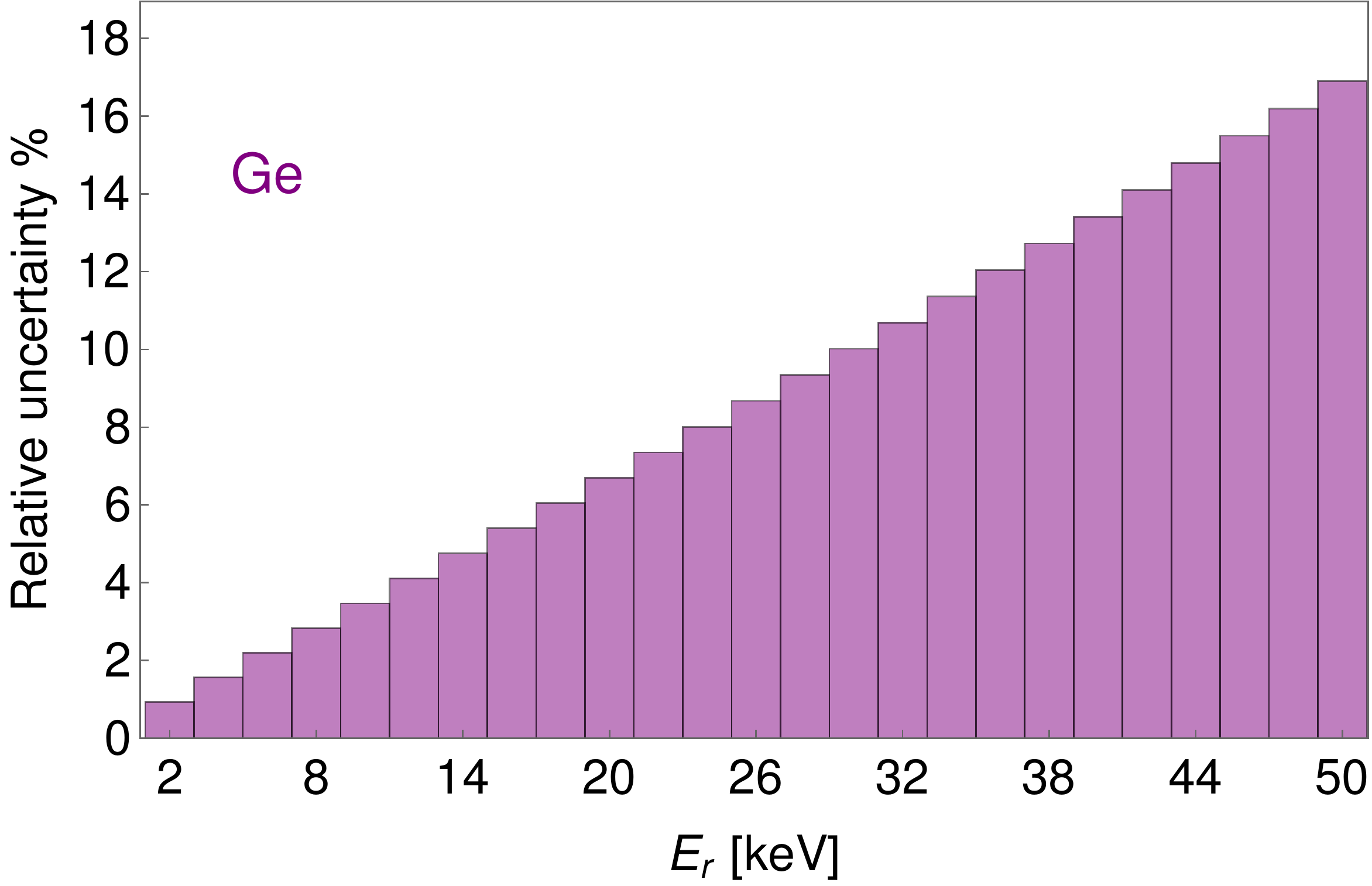}\\
\hfill
 \includegraphics[scale=0.37]{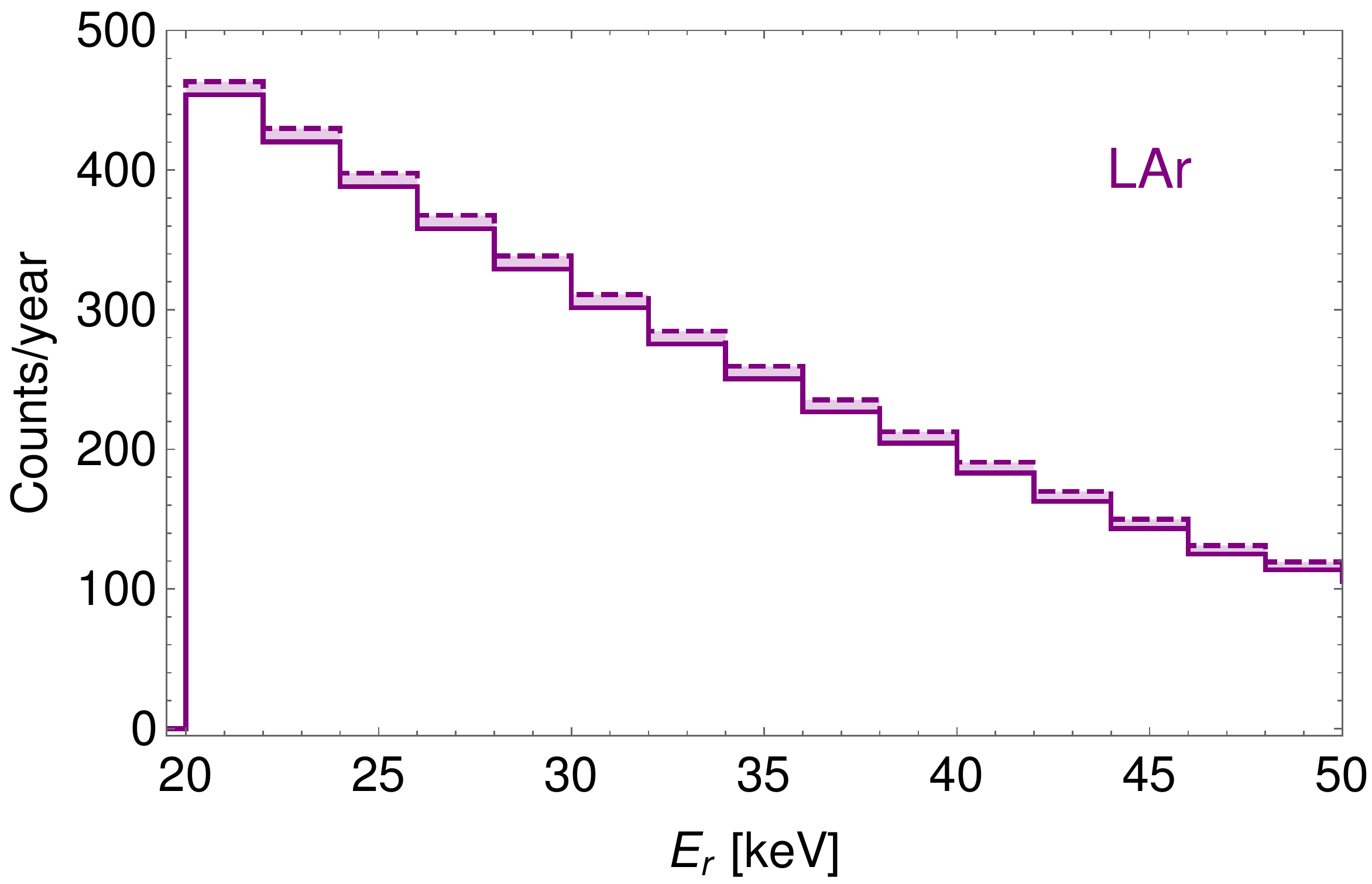}
  \hfill
  \includegraphics[scale=0.353]{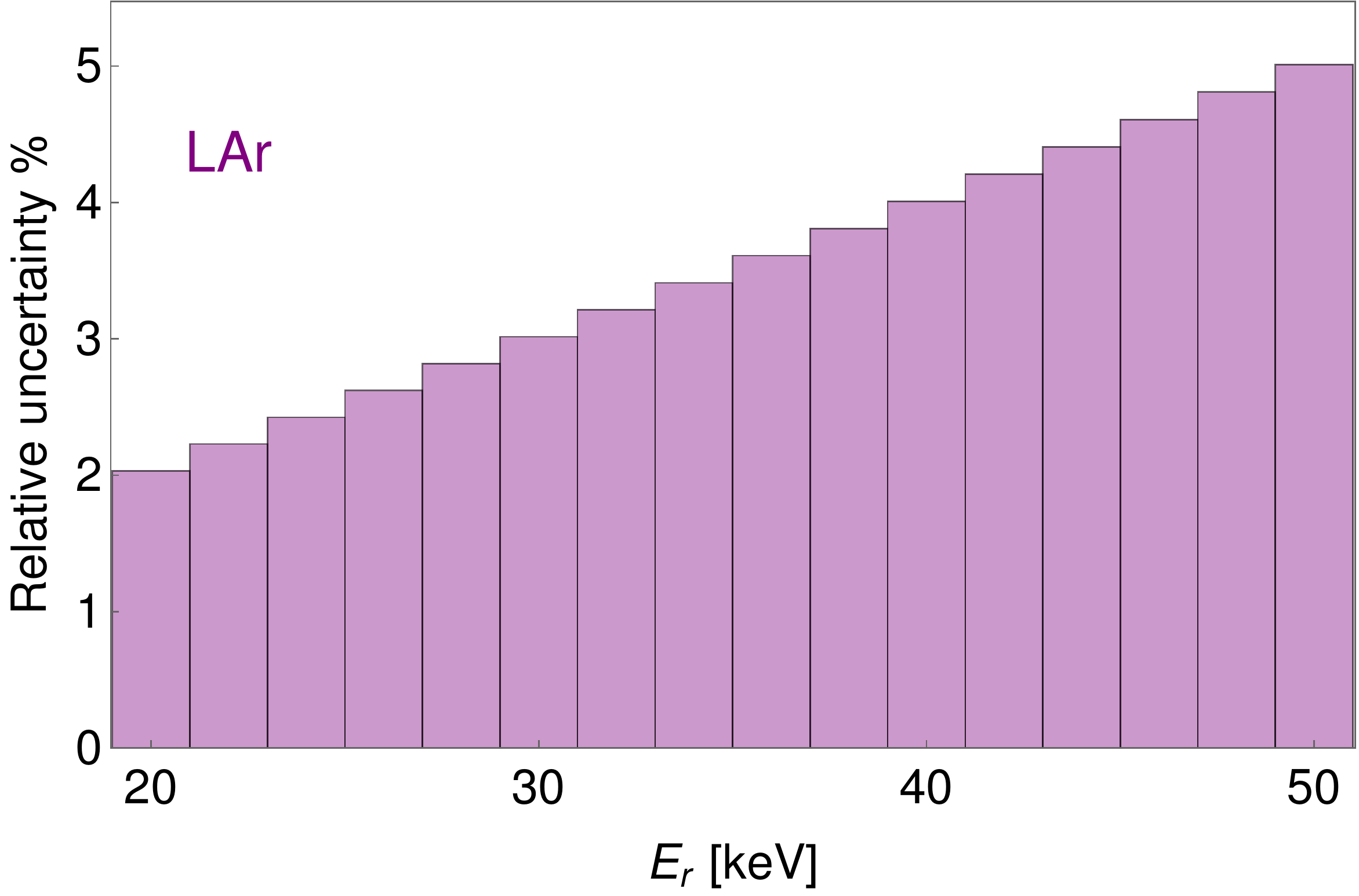}\\
  \hfill
  \includegraphics[scale=0.37]{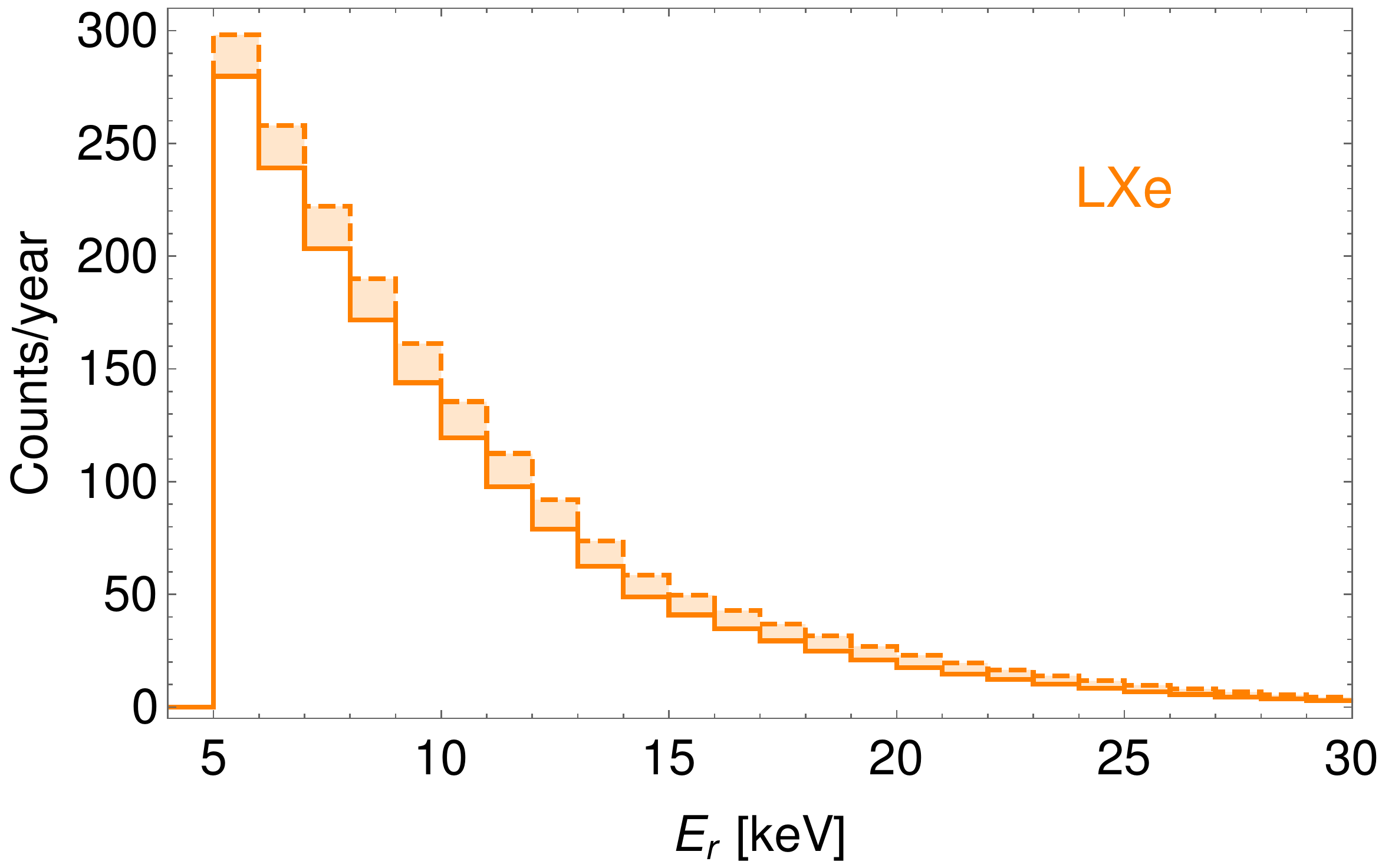}
  \hfill
   \includegraphics[scale=0.36]{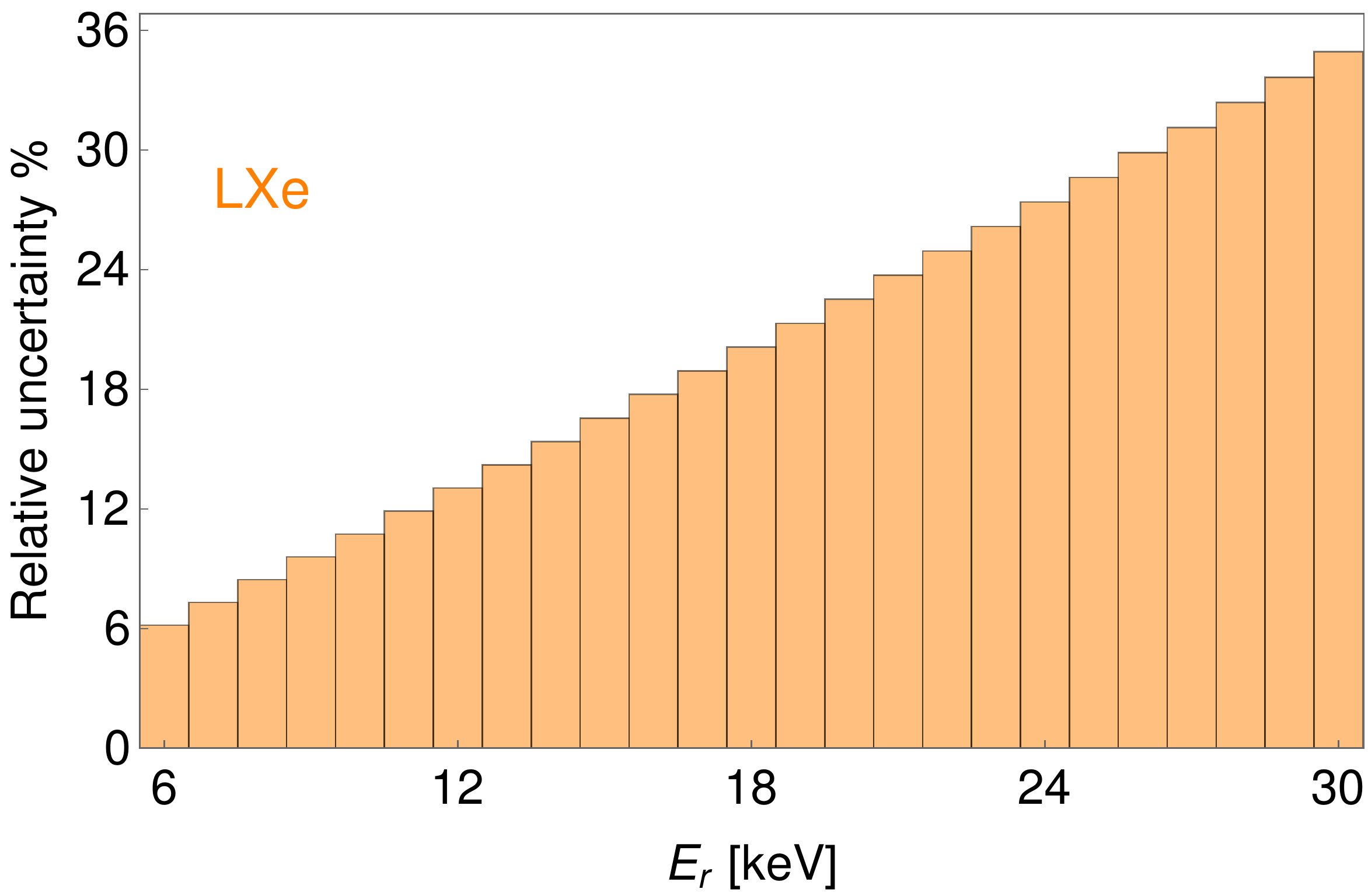}
  \caption{Expected number of CE$\nu$NS events (\textbf{left}) and its relative uncertainty, equal to (maximum count number $-$ minimum count number)$/$(maximum count number)$\times 100$,  (\textbf{right})
   as a function of recoil energy for argon, germanium  and xenon. The calculations have been done for the
    COHERENT Ge, LAr and LXe detectors including form factor uncertainties,
    summing the three neutrino flavor contributions and taking
    $n_\text{POT}$ as in the CsI case.}
  \label{fig:recoil-spectra}
\end{figure*}

COHERENT phase-II consists of a p-type point-contact high purity
germanium detector with $m_\text{det}=15$~kg and located at $L=22$~m
from the source. COHERENT phase-III, instead, aims at measuring
CE$\nu$NS by using a two-phase liquid xenon detector with
$m_\text{det}=100$~kg and located at $L=29$~m. A one ton LAr detector at $L=29$~m
is also under consideration. At low recoil energies
the number of CE$\nu$NS events in the Xe detector will exceed those in
the Ge detector by about an order of magnitude. However, since Ge
isotopes are lighter than Xe isotopes, the Ge detector will be
sensitive to CE$\nu$NS events at higher recoil energies and so they
are complementary~\cite{Akimov:2015nza}. Using these target masses,
the corresponding locations and assuming $n_\text{POT}$/year as in the
CsI calculation, we calculate the impact of the uncertainties on the
expected number of events in both detectors.   As can be seen from
  Fig.~\ref{fig:recoil-spectra} in the germanium case relative
  uncertainties can be sizable, and for Xe, relative
  uncertainties are still larger. It is clear that form factor uncertainties
 should be taken into account in the analysis of COHERENT data.

Since form factor uncertainties increase with increasing momentum
transfer, they are also relevant for CE$\nu$NS induced by the DSNB and
sub-GeV atmospheric neutrinos. DSNB neutrinos (neutrinos and
antineutrinos of all flavors) result from the cumulative emission from
all past core-collapse supernovae. Their flux is thus determined by
the rate for core-collapse supernova (determined in turn by the cosmic
star formation history), and the neutrino emission per supernova,
properly redshifted over cosmic time~\cite{Ando:2004hc} (see
Appendix~\ref{sec:DSNB} for details).  The latter is well described by
a Fermi-Dirac distribution with zero chemical potential and with
$T_{\nu_e}<T_{\bar\nu_e}<T_{\nu_x}$ \cite{Horiuchi:2008jz}. For the
calculation of the DSNB neutrino flux we use $T_{\nu_e}=3$~MeV,
$T_{\bar\nu_e}=5$~MeV and $T_{\nu_x}=8$~MeV, and sum over all flavors.

Atmospheric neutrino fluxes ($\nu_e$ and $\nu_\mu$ and their
antiparticles) result from hadronic showers induced by cosmic rays in
the Earth's atmosphere. We take the atmospheric fluxes from
Ref.~\cite{Battistoni:2005pd} generated by a FLUKA Monte Carlo
simulation~\cite{Ferrari:2005zk}, that includes $\nu_{e,\tau}$ and
$\bar\nu_{e,\tau}$ fluxes up to about $10^3$~MeV.  We only consider
atmospheric neutrino fluxes below 100~MeV because for higher energies
the loss of coherence for CE$\nu$NS drastically depletes the neutrino
event rate making the flux at those energies less relevant.

\begin{figure*}
  \centering
  \includegraphics[scale=0.37]{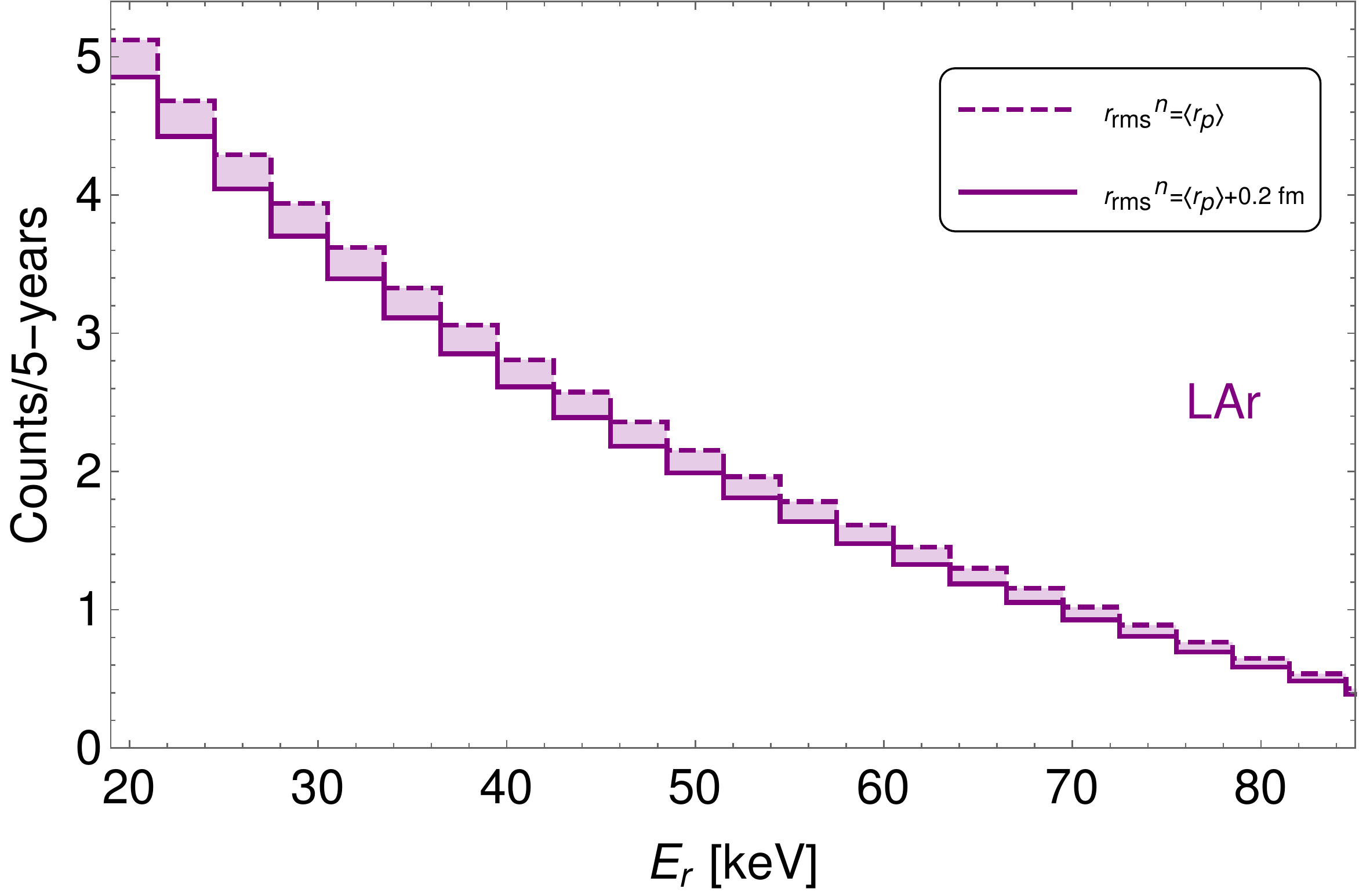}
  \includegraphics[scale=0.378]{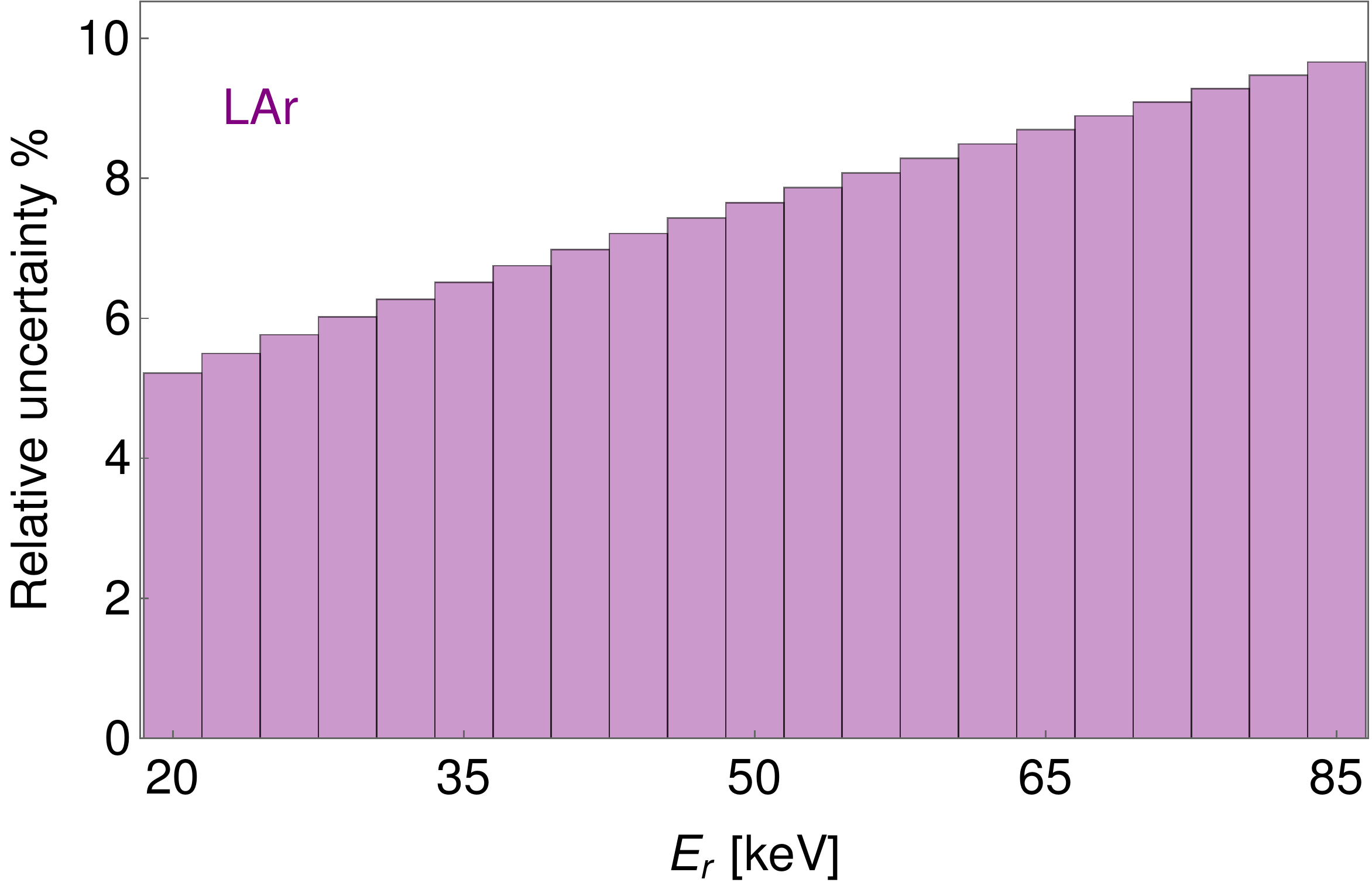}
  \caption{
  Recoil spectrum (\textbf{left}) and its relative uncertainty (\textbf{right}) from DSNB and atmospheric neutrinos at an Ar-based dark matter detector assuming an exposure of 1000 ton-year, as expected for Argo.
       The dashed (solid) histrograms correspond
      to $r_\text{rms}^n=r_\text{rms}^p$
      ($r_\text{rms}^n=r_\text{rms}^p+0.2$~fm). The DSNB contribution is subdominant but not negligible.}
  \label{fig:dsnb-atm-fig}
\end{figure*}

Figure~\ref{fig:dsnb-atm-fig} shows the event spectrum
  for the sum of the DSNB and atmospheric neutrino
  contributions in an argon detector with an exposure of $1000$~ton-year.
   The dashed (solid) histogram is obtained by fixing
  $r_\text{rms}^n=r_\text{rms}^p$ 
  ($r_\text{rms}^n=r_\text{rms}^p+0.2$~fm). In the calculation we
  include only $^{40}\text{Ar}$ and checked that form factor
  uncertainties for the ``high energy'' tail of the solar neutrino
  spectrum ($^8$B and hep neutrinos) are not relevant, as expected
  from the middle panel in
  Fig.~\ref{fig:Helm-vs-q-and-percentage-uncertainty}. The
  DSNB flux dominates in the window $\sim 18-32$~MeV, just above the
  kinematic tail of
  hep neutrino spectrum. Since the DSNB flux dominates
  only in that narrow window its contribution to the total event
  rate spectrum is subdominant, but sizable enough to contribute
  to the event rate spectrum. The relative uncertainty in the lowest energy bin is 5\% and gets
  larger for larger recoil energies.
  
\section{Implications for new physics searches}
\label{sec:new-physics}
We now discuss the effects of the neutron form factor uncertainties on
the predictions for new physics. To do so, we consider three new
physics scenarios that have been discussed in the literature in
connection with COHERENT data: NSI~\cite{Liao:2017uzy,Kosmas:2017tsq},
sterile neutrinos~\cite{Kosmas:2017tsq} and NGI~\cite{AristizabalSierra:2018eqm}.
\subsection{Theoretical basics}
\label{sec:theory-basics}
NSI is a parameterization of a new physics neutral current interaction
mediated by a vector boson of mass $m_V$~\cite{Wolfenstein:1977ue}.
Dropping the axial coupling, which yields nuclear spin-suppressed
effects, and in the limit $m_V\gg q_\text{CEvNS}$,
\begin{equation}
  \label{eq:NSI-Lag}
  \mathcal{L}_\text{NSI}=\frac{G_F}{\sqrt{2}}\sum_{q=u,d}
  \left[\overline{\nu}_i\,\gamma_\mu\,(1-\gamma_5)\,\nu_j\right]\,
  \left[\overline{q}\,\gamma^\mu\epsilon_{ij}^q\,q\right]\ .
\end{equation}
Written this way, the NSI parameters measure the strength of the new
interaction compared to the weak interaction,
$\epsilon_{ij}^q\simeq g_{qij}^2/m_V^2/G_F$, where $g_{qij}$ are gauge
couplings. In the presence of NSI, the differential cross section
becomes lepton-flavor dependent. For the $i^{\rm{th}}$ neutrino flavor
it can be derived from Eq.~(\ref{eq:cevns-xsec}) by trading
$g_V^n\to g_V^n+\epsilon_{ij}^n$ and $g_V^p\to g_V^p+\epsilon_{ij}^p$,
with $\epsilon_{ij}^n=\epsilon_{ij}^u+2\epsilon_{ij}^d$ and
$\epsilon_{ij}^p=2\epsilon_{ij}^u+\epsilon_{ij}^d$~\cite{Barranco:2005yy,Scholberg:2005qs}.

Oscillations with an eV mass sterile neutrino have an effect on the
CE$\nu$NS event rate. If the flux of $\nu_\alpha$ neutrinos
($\alpha=e, \mu, \tau$) at the source is $\Phi_{\nu_\alpha}(E_\nu)$,
the flux at the detector will be diminished by the fraction of
neutrinos that oscillate into the sterile and the other active
states. Quantitatively this means that the flux of neutrinos of flavor
$\alpha$ that reach the detector is
$\mathcal{P}_{\alpha\alpha}\Phi_{\nu_\alpha}(E_\nu)$, where
$\mathcal{P}_{\alpha\alpha}$ is the $\nu_\alpha$ survival probability
defined as
$\mathcal{P}_{\alpha\alpha}=1-\mathcal{P}_{\alpha
  s}-\mathcal{P}_{\alpha\beta}$
$(\alpha\neq\beta)$, with $\mathcal{P}_{\alpha s}$ and
$\mathcal{P}_{\alpha\beta}$ the $\nu_\alpha\to \nu_s$ and
$\nu_\alpha\to \nu_\beta$ neutrino oscillation probabilities,
respectively. For short-baseline experiments $\mathcal{P}_{\alpha s}$
is given by
\begin{equation}
  \label{eq:survival-prob}
  \mathcal{P}_{\alpha s}=\sin^22\theta_{\alpha\alpha}\,
  \sin^2
  \left[1.27
    \left(
      \frac{\Delta m_{41}^2}{\text{eV}^2}
    \right)
    \left(\frac{L}{\text{m}}\right)
    \left(\frac{\text{MeV}}{E_\nu}\right)
  \right]\ .
\end{equation}
Here,
$\sin^22\theta_{\alpha\alpha}=4|U_{\alpha 4}|^2(1-|U_{\alpha 4}|^2)$
($U$ is the $4\times 4$ lepton mixing matrix) and
$\Delta m_{41}^2=m_4^2-m_1^2$ is the sterile-active neutrino
mass-squared difference. The oscillation probability for active
states is
\begin{equation}
  \label{eq:active-oscillation-prob}
  \mathcal{P}_{\alpha\beta}=\sin^4\theta_{\alpha\beta}\sin^2
  \left[1.27
    \left(
      \frac{\Delta m_{41}^2}{\text{eV}^2}
    \right)
    \left(\frac{L}{\text{m}}\right)
    \left(\frac{\text{MeV}}{E_\nu}\right)
  \right]\ ,
\end{equation}
where $\sin^4\theta_{\alpha\beta}=4|U_{\alpha 4}|^2|U_{\beta 4}|^2$.
The recoil spectrum induced by neutrinos of flavor $\alpha$ is then
given by
\begin{equation}
  \label{eq:recoil-spectrum-sterile-nu}
  \frac{dR_{\nu_\alpha}}{dE_r}=N_T\sum_\beta \int dE_\nu
  \left[
    (1 - \mathcal{P}_{\alpha s}- \mathcal{P}_{\alpha\beta})\Phi_{\nu_\alpha}
    +\mathcal{P}_{\alpha\beta}\Phi_{\nu_\beta}
  \right]\frac{d\sigma}{dE_r}\,.
\end{equation}
To a fairly good approximation
$\mathcal{P}_{\alpha\beta}$ can be neglected due to the higher-order
active-sterile mixing suppression.  $N_T$ is the number
of target nuclei  and $d\sigma/dE_r$ is the SM cross
section in Eq.~(\ref{eq:cevns-xsec}).  The total number of counts in the
$k^{\rm{th}}$ bin is obtained from Eq.~(\ref{eq:recoil-spectrum-sterile-nu})
according to
\begin{equation}
  \label{eq:total-num-events-sterile-neu}
  R=\sum_\alpha R_{\nu_\alpha} = \sum_\alpha \int_{E_k-\Delta E_k}^{E_k+\Delta E_k}
  \,\mathcal{A}(E_r)\,\frac{dR_{\nu_\alpha}}{dE_r}dE_r\ .
\end{equation}
Experimental information on $R$ can then be mapped into
$\sin^2\theta_{ij}-\Delta m_{41}^2$ planes.

NGI follows the same approach as NSI, but includes all possible
Lorentz-invariant structures~\cite{Lee:1957qr}.  It was
introduced in the analysis of neutrino propagation in matter in Ref.~\cite{Bergmann:1999rz}, 
studied in the context of CE$\nu$NS physics in Ref.~\cite{Lindner:2016wff} and in the light of COHERENT data in
Ref.~\cite{AristizabalSierra:2018eqm}.  Dropping flavor indices, the most
general Lagrangian reads
\begin{equation}
  \label{eq:NGI-Lag}
  \mathcal{L}_\text{NGI}=\frac{G_F}{\sqrt{2}}\sum_{\substack{a=S,P,V,A,T\\q=u,d}}
  \left[\overline{\nu}\,\Gamma^a\,\nu\right]\;
  \left[\overline{q}\,\Gamma_a\left(C_a^q + i\gamma_5\,D_a^q\right)q\right]\ ,
\end{equation}
where
$\Gamma_a=\{\mathbb{I},i\gamma_5,\gamma_\mu,\gamma_\mu\gamma_5,\sigma_{\mu\nu}\}$,
with $\sigma_{\mu\nu}=i[\gamma_\mu,\gamma_\nu]/2$. As in the NSI case,
some of these couplings lead to spin-suppressed interactions which we
do not consider. Relevant couplings therefore include all Lorentz
structures for the neutrino bilinear and only scalar, vector and
tensor structures for the quark currents. For the NGI analysis, we consider only
one Lorentz structure at a time and assume the $C$ and $D$ parameters to be real. 
We may therefore consider the individual cross sections.
Assuming a spin-1/2 nuclear ground state and neglecting
$\mathcal{O}(E_r^2/E_\nu^2)$ terms, 
\begin{align}
  \label{eq:xsecs-NGI}
  \frac{d\sigma_S}{dE_r}&=\frac{G_F^2m_N}{8\pi}\xi_S^2(q^2)
  \frac{E_rm_N}{E_\nu^2}\ ,
  \nonumber\\
  \frac{d\sigma_V}{dE_r}&=\frac{G_F^2m_N}{8\pi}\xi_V^2(q^2)
  \left(2-\frac{E_rm_N}{E_\nu^2}-\frac{2E_r}{E_\nu} \right)\ ,
  \nonumber\\
  \frac{d\sigma_T}{dE_r}&=\frac{G_F^2m_N}{2\pi}\xi_T^2(q^2)
  \left(2-\frac{E_rm_N}{2E_\nu^2}-\frac{2E_r}{E_\nu}\right)\ .
\end{align}
\begin{figure*}
  \centering
  \includegraphics[scale=0.37]{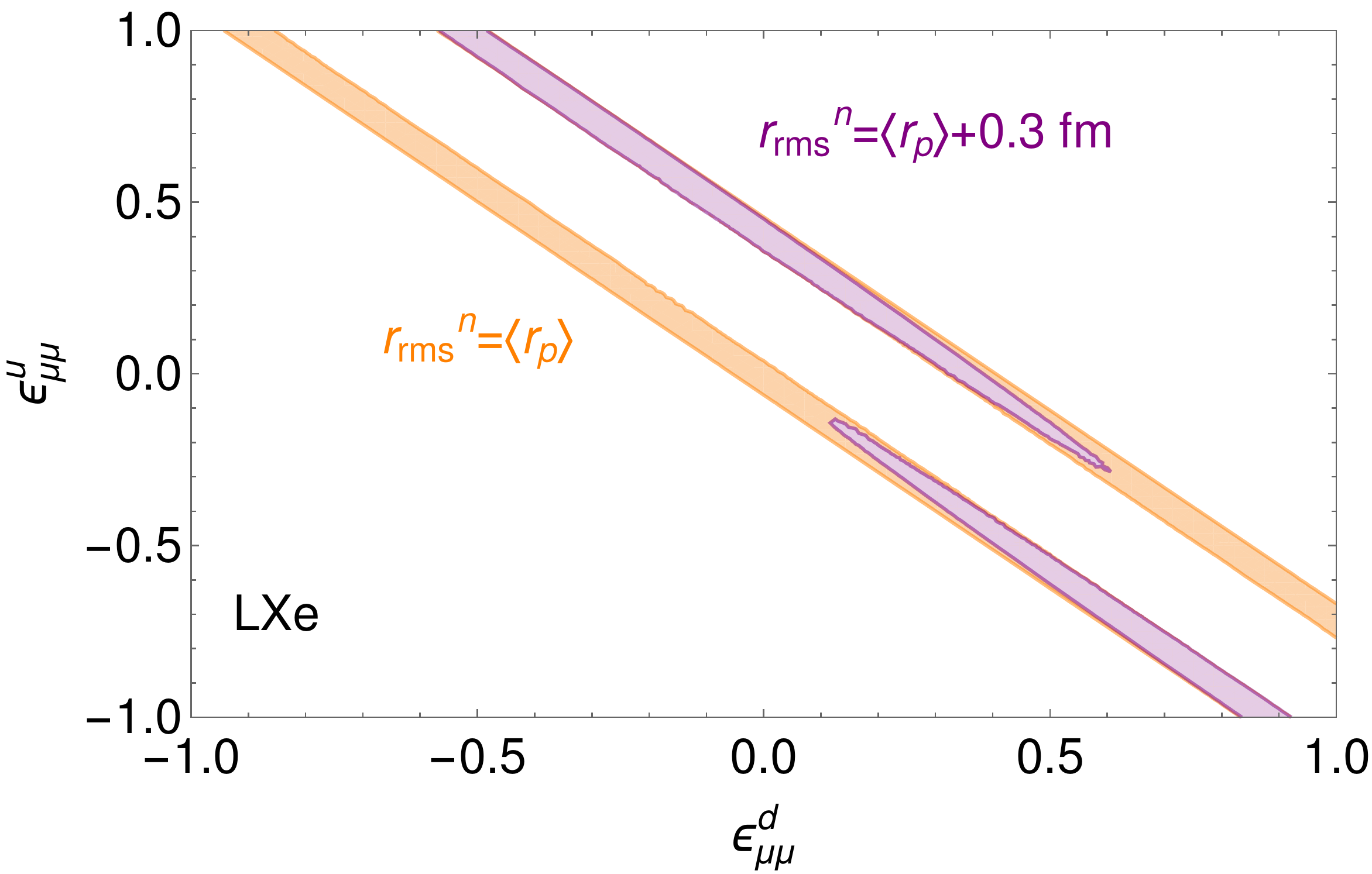}
  \hfill
  \includegraphics[scale=0.37]{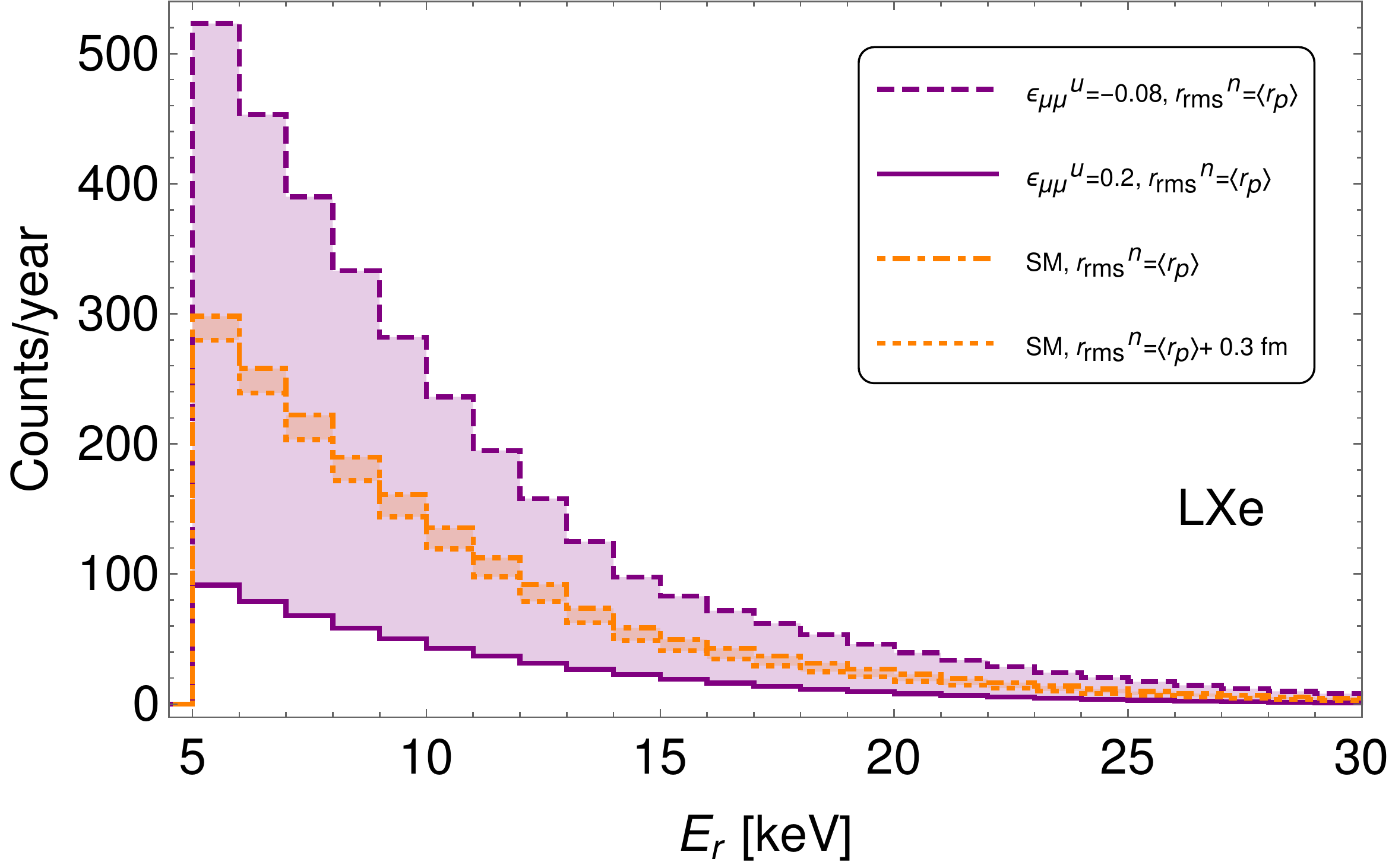}
  \vspace{0.4cm}\\
  \caption{\textbf{Left}: 90\% C.L. allowed regions in the NSI case
    and for two choices of the rms neutron radius.  The diagonal bands
    are obtained assuming $r_\text{rms}^n=\langle r_p\rangle=4.79$~fm
    for all xenon isotopes. The purple regions are obtained for
    $r_\text{rms}^n=5.09$~fm (for all
    isotopes). \textbf{Right}: Number of events as a function of
    recoil energy for $\epsilon_{\mu\mu}^u\subset [-0.088,0.37]$
    \cite{Coloma:2017ncl} and all other couplings equal to zero
    (purple histograms). The orange histograms correspond to the SM
    prediction including form factor uncertainties.}
  \label{fig:nsi-and-sterile}
\end{figure*}
For the scalar and tensor cases the SM cross section has to be added. In the
vector case $\xi_V^2$ includes the SM contribution which interferes
with the BSM vector piece. The definition of the different parameters
in Eq.~(\ref{eq:xsecs-NGI}) can be found in Appendix~\ref{sec:parameters-def}.
\subsection{Impact of neutron form factor uncertainties}
\label{sec:uncertainties-ff-BSM}
We calculate the impact of uncertainties in the neutron rms radii for
CsI, Ge and Xe in the presence of NSI. To do so, we take as
``experimental'' input the number of events predicted by the SM
assuming
$r_\text{rms}^n=\langle r_p\rangle=\sum_i
r_\text{rms}^{p\,i}X_i=4.06$~fm
for all germanium isotopes and $4.79$~fm for all xenon isotopes. Here
$r_\text{rms}^{p\,i}$ is the rms radius of the proton distribution of the
$i^{\rm{th}}$ isotope with abundance $X_i$.  We proceed
as we have done in Section~\ref{sec:implications-1}, i.e., for CsI we take
into account the Cs and I contributions, while for Ge and Xe the
contributions for each isotope according to
Eq.~(\ref{eq:recoil-spectrum}). For all three cases we
  assume four years of data taking. For our analysis we define the
$\chi^2$ function,
\begin{equation}
  \label{eq:chi-fun}
  \chi^2=\sum_{i}
  \left(\frac{N^\text{meas}_i-(1+\alpha)N^\text{BSM}_i(\mathcal{P})}{\sigma_i}\right)^2
  + \left(\frac{\alpha}{\sigma_\alpha}\right)^2\ ,
\end{equation}
where $\alpha$ is a nuisance parameter that accounts for uncertainties
in the signal rate, $N^\text{meas}_i$ is the number of simulated
events in the $i^{\rm{th}}$ bin, and $N^\text{BSM}_i$ is the number of
predicted events in the BSM scenario (which depend on the set of
parameters $\mathcal{P}$). The statistical uncertainty
  in the simulated data is
  $\sigma_i=\sqrt{N_i^\text{meas}+B_i}$, where
  $B_i$ includes the beam-on and twice the steady-state neutron
  background. Beam-on neutrons are neutrons from the spallation source that penetrate the 19.3~m
  of moderating material, and steady-state neutrons are produced by cosmic rays  interacting with the shielding material and by
  radioactivity. We select a 5~keV analysis threshold for the Ge and LXe detectors so that the neutron background can be assumed to be flat. It is anticipated that the shielding structures for these detectors will reduce the background rate well below the SM 
  CE$\nu$NS expectation~\cite{Akimov:2018ghi}. With that in mind, we set $\sum_i B_i$  equal to 50\% of the
  SM signal between $5-30$~keV for the Ge and LXe detectors; this implies that the total steady-state background
  between $5-30$~keV is approximately 25\% of the SM signal. In the future, the quenching factor 
  uncertainty is expected to be reduced to $12.5\%$~\cite{Grayson}. Keeping the neutrino flux and signal
  acceptance uncertainties unchanged from their current values, i.e., $10\%$ and $5\%$, respectively, we have
   the systematic uncertainty $\sigma_\alpha=0.168$. Our simplification that the systematic uncertainty is correlated between 
   bins is unavoidable given publicly available information. 
   
Assuming $\epsilon_{\mu\mu}^q\subset [-1,1]$ we determine the 90\%~C.L. exclusion regions in two cases,
$r_\text{rms}^n=\langle r_p\rangle$ and
$r_\text{rms}^n=\langle r_p\rangle + 0.3$~fm. We find
that the CsI and Ge detectors are rather insensitive to the choice of
the neutron rms radius; the resulting 90\%~C.L. regions barely change
with $r_\text{rms}^n$. For Xe, the result is quite different. Changing
$r_\text{rms}^n$ has a strong impact on the available parameter
space. This can be seen from the left panel of
Fig.~\ref{fig:nsi-and-sterile}, where the diagonal bands are obtained
in the case $r_\text{rms}^n=\langle r_p\rangle$, while the purple
regions are obtained for
$r_\text{rms}^n=5.09$~fm. This
result is as expected. Firstly, the LXe detector has a larger target
mass (about a factor 6.5 larger compared to the CsI and Ge detectors),
so for a common data taking time the accumulated statistics in the LXe
detector is larger. Secondly, the number of events expected in the NSI
scenario with $r_\text{rms}^n=\langle r_p\rangle$ reproduces the
simulated data better than with
$r_\text{rms}^n= 5.09$~fm since the data are
simulated with $r_\text{rms}^n=\langle r_p\rangle$.  For the rest of our NSI study
we only consider a large LXe detector. Note, however, that
increasing the exposure for the CsI or Ge detectors will change the
situation. In doing so these detectors will become---as the LXe
detector---sensitive to uncertainties in the neutron rms radii.
On the other hand, the corresponding results for a large LAr detector are not qualitatively affected
by form factor uncertainties.

\begin{figure*}
  \centering
  \includegraphics[scale=0.37]{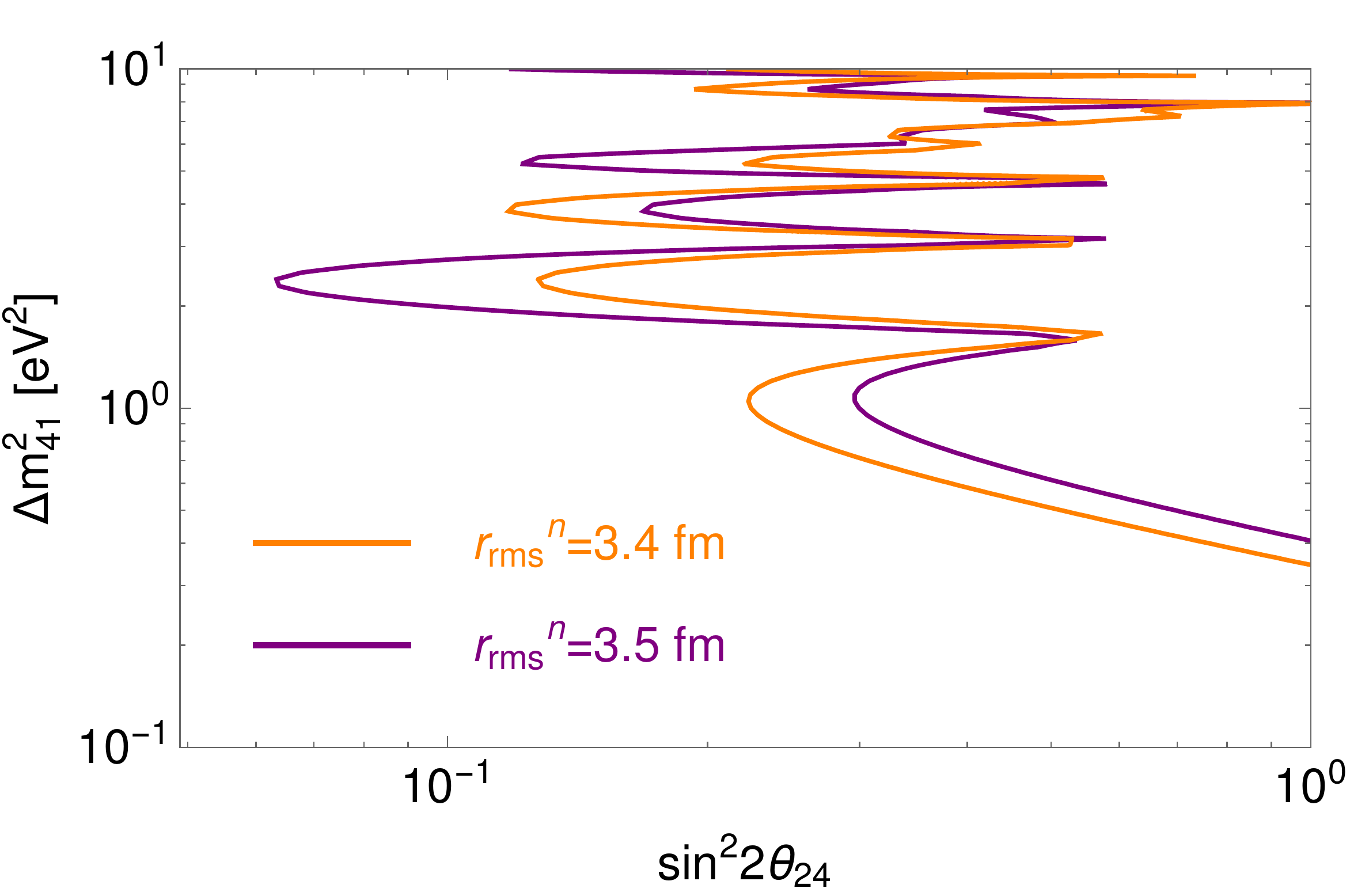}
  \hfill
  \includegraphics[scale=0.37]{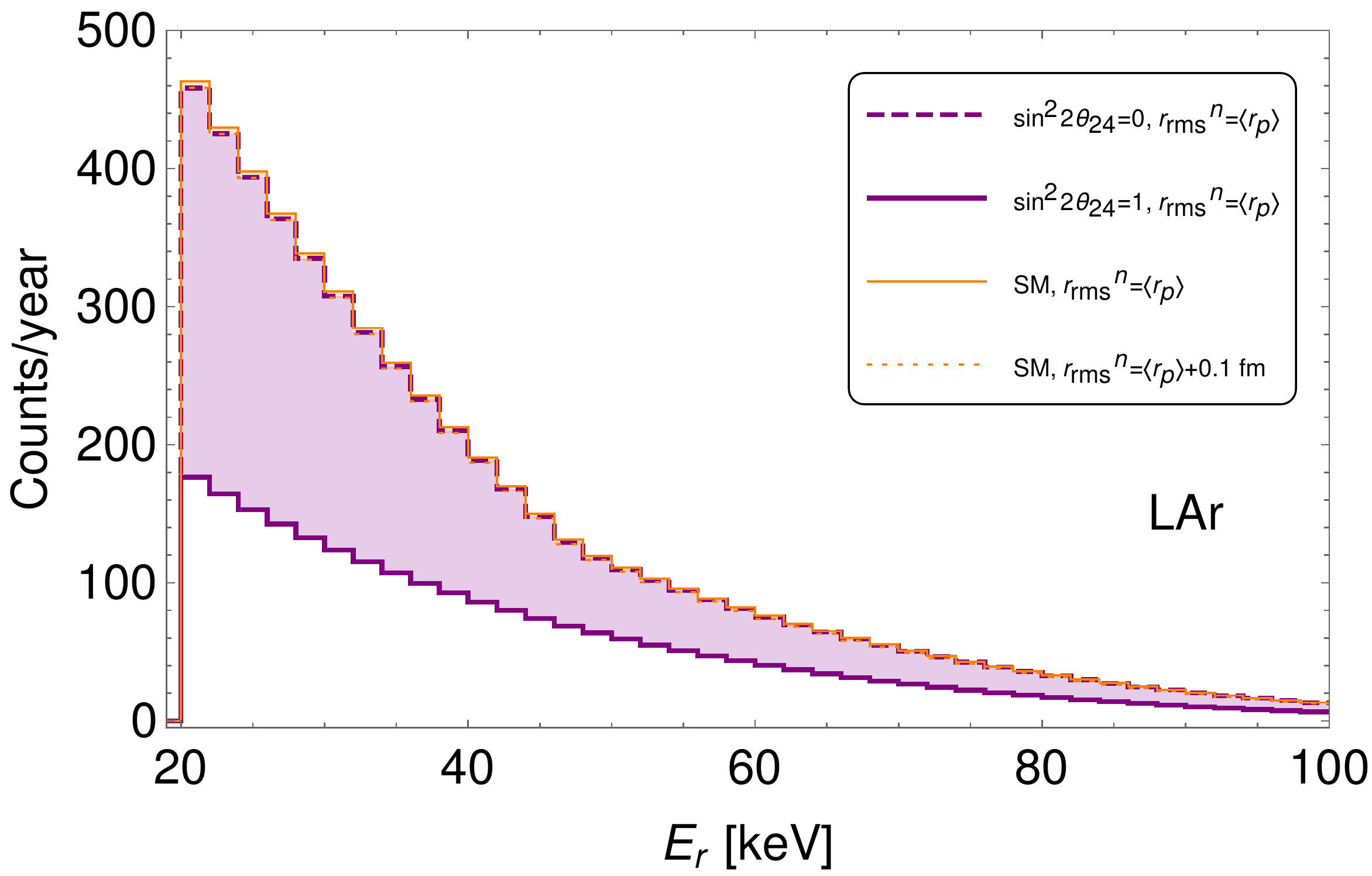}
  \caption{\textbf{Left}: 90\% C.L. exclusion regions in the case of
    sterile neutrinos in the 3+1 scheme obtainable from a
      one ton LAr COHERENT detector with four years of data taking. The neutron rms
    radius is $\langle r_p\rangle$ for the orange contour and
    3.5~fm for the purple contour. \textbf{Right}:
    The orange histograms show the SM expectation for the event
    spectrum including neutron form factor uncertainties, while the
    purple histograms are the spectra expected by fixing
    $r_\text{rms}^n=\langle r_p\rangle=3.4$~fm,
    $\Delta m_{41}^2=1.3$~eV$^2$, $\sin^2\theta_{14}=0.01$,
    $\theta_{34}=0$, with $\sin^2 2\theta_{24}=0$ and $1$.}
  \label{fig:sterile-nu-fig}
\end{figure*}
To determine the extent to which neutrino NSI can be distinguished
from the SM signal including its neutron form factor uncertainties, we
calculate the number of events assuming $\epsilon_{\mu\mu}^d=0$ and
$\epsilon_{\mu\mu}^u\subset [-0.088,0.37]$. These values correspond to
the 90\% C.L. range obtained from global fits to neutrino oscillation
data including COHERENT (CsI phase) data without accounting for
energy-dependent form factor uncertainties~\cite{Coloma:2017ncl}.  The result is shown in the right panel
of Fig.~\ref{fig:nsi-and-sterile}. The NSI (purple) histograms are
obtained by fixing $r_\text{rms}^n=\langle r_p\rangle$.  The SM
histograms (orange) are instead obtained by fixing
$r_\text{rms}^n=\langle r_p\rangle$ (upper boundary) and
$r_\text{rms}^n=5.09$~fm (lower boundary) and
determines the SM expectation within the form factor uncertainties.
Clearly, the SM expectation with form factor uncertainties lies within
the NSI expectation for $\epsilon_{\mu\mu}^u$ between $-0.08$ and 0.2
with a fixed form factor.  There are various ranges of NSI couplings
that will produce signals that cannot be disentangled from the SM
signal. This will persist unless uncertainties on the neutron rms
radii are reduced. We have chosen $\epsilon_{\mu\mu}^u$ to stress this
point, although results for $\epsilon_{\mu\mu}^d$, $\epsilon_{ee}^q$,
$\epsilon_{\tau\tau}^q$ and $\epsilon_{e\tau}^q$, will lead to the
same conclusion.  Needless to say, allowing for multiple nonzero NSI
parameters will further complicate the ability to discriminate new
physics from the SM.

For sterile neutrinos we display the 90\% C.L. exclusion regions in
the $\sin^22\theta_{24}-\Delta m_{41}^2$
plane. We highlight the exquisite sensitivity required to probe $3+1$ oscillations
by assessing the capability of a future one-ton LAr COHERENT detector which has the advantage of smaller
form factor uncertainties.  The results for four years of data taking
are shown in the left panel of Fig.~\ref{fig:sterile-nu-fig}. The analysis is similar to that for the LXe
detector except that here we set $\sum_i B_i$  equal to 50\% of the
  SM signal between $20-100$~keV. The
contours are obtained for $N^\text{BSM}$ calculated for
$r_\text{rms}^n=\langle r_p\rangle=3.4$~fm (orange contour) and
$r_\text{rms}^n=3.5$~fm (purple contour). We fix
$\sin^2\theta_{14}=0.01$ (best-fit value from a global fit to $\nu_e$
and $\bar{\nu}_e$ disappearance data~\cite{Dentler:2018sju}) and
$\theta_{34}=0$. This result demonstrates that the available regions
in parameter space have a strong dependence on the neutron rms
radii. A $\sim 3\%$ change in $r_\text{rms}^n$ is sufficient to
significantly modify the results of the parameter fit.  

It is clear that a more precise treatment of sterile neutrino effects
should include neutron form factor uncertainties, otherwise one might
end up misidentifying SM uncertainties with these effects. To show
this might be the case, we calculate the number of events for sterile
neutrino parameters fixed as in the previous calculation and for
$\Delta m_{41}^2=1.3$~eV$^2$,
$r_\text{rms}^n=\langle r_p\rangle$, $\theta_{24}=0$ and
$\sin^22\theta_{24}=1$. We then compare the resulting (purple) histograms 
 with the SM predictions including uncertainties (in
orange); see the right panel of Fig.~\ref{fig:sterile-nu-fig}.
The overlapping spectra show that
an identification of the new effects is not readily possible.

\begin{figure*}
  \centering
  \includegraphics[scale=0.357]{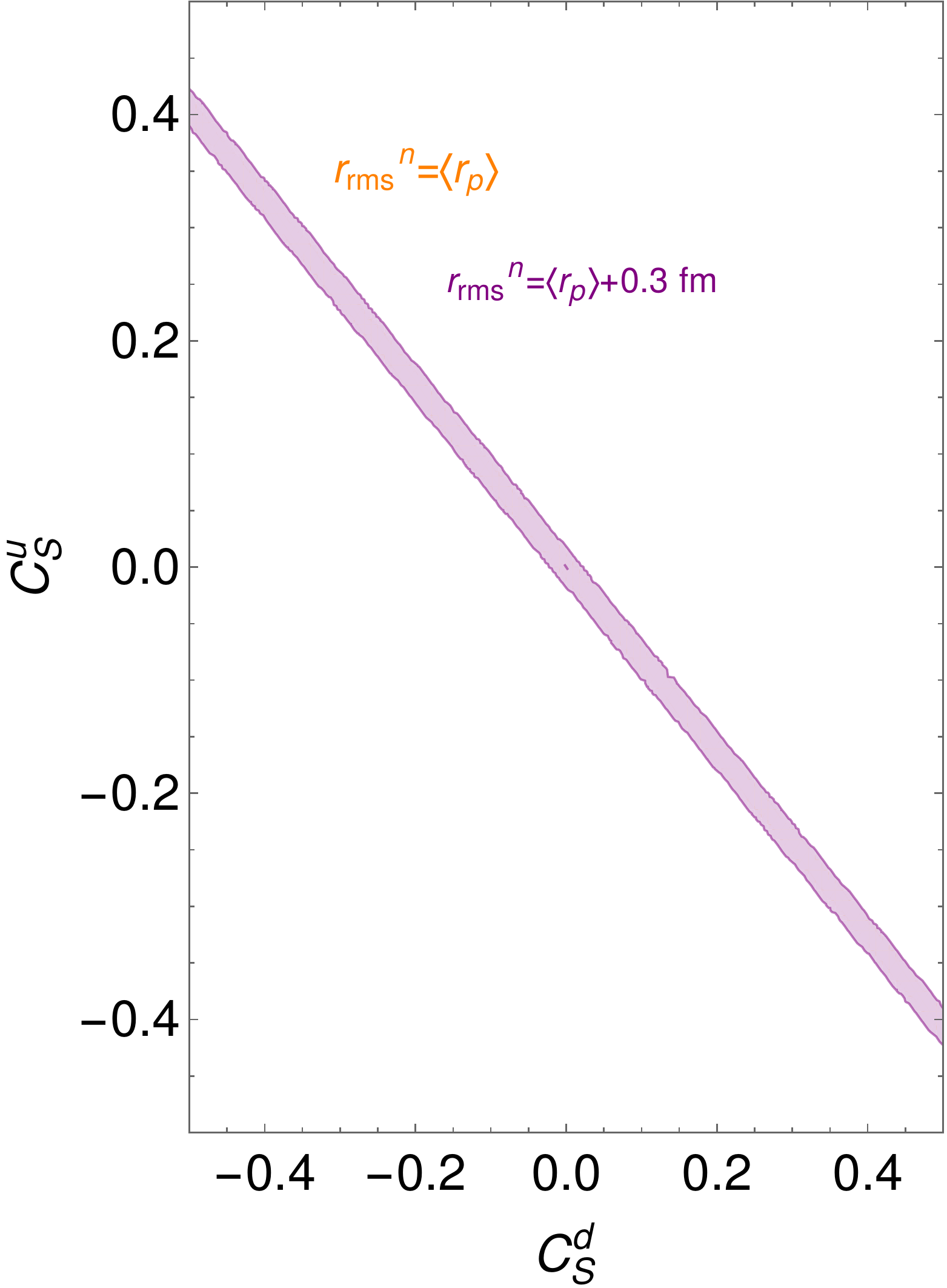}
  \includegraphics[scale=0.37]{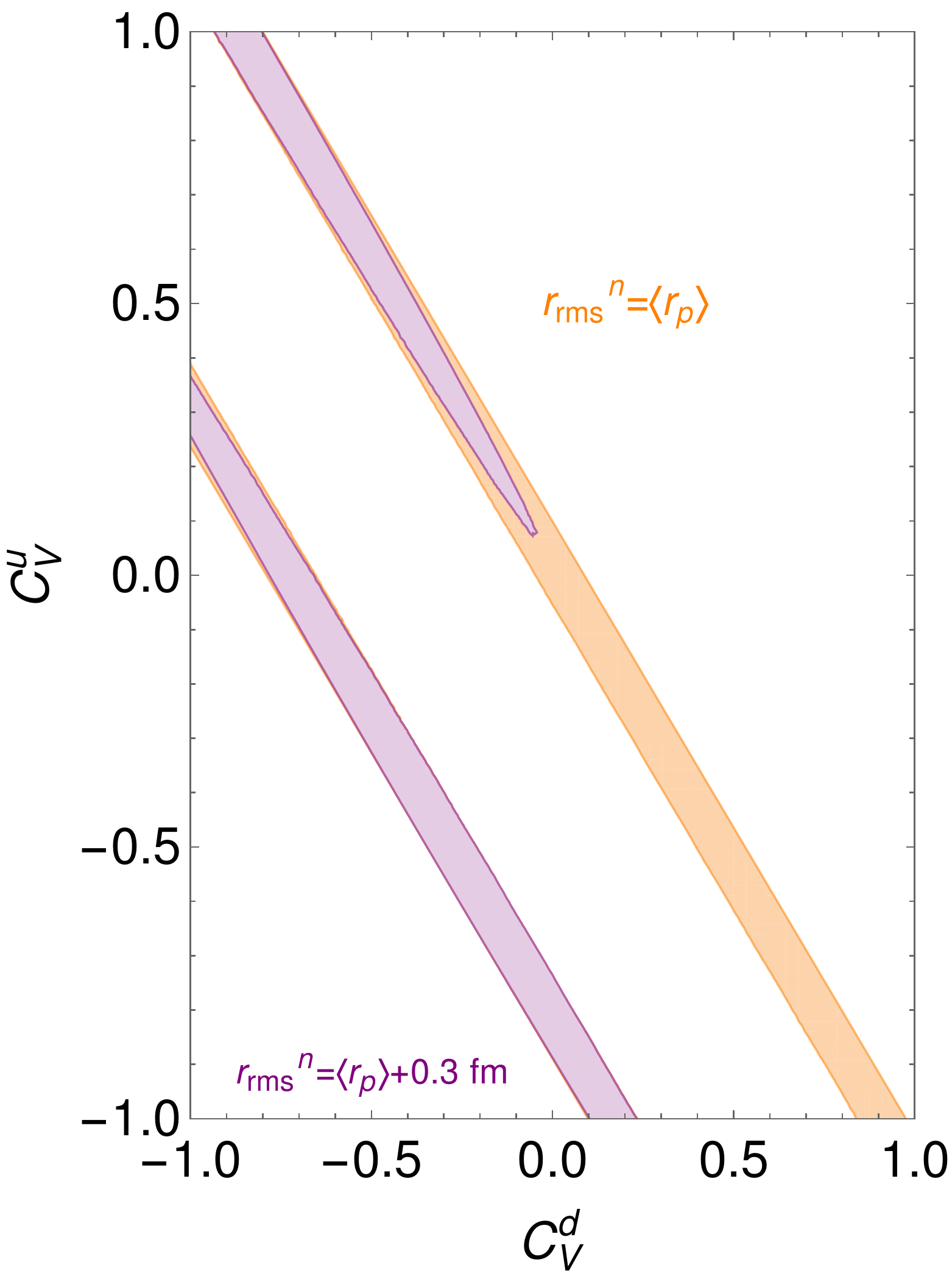}
  \includegraphics[scale=0.37]{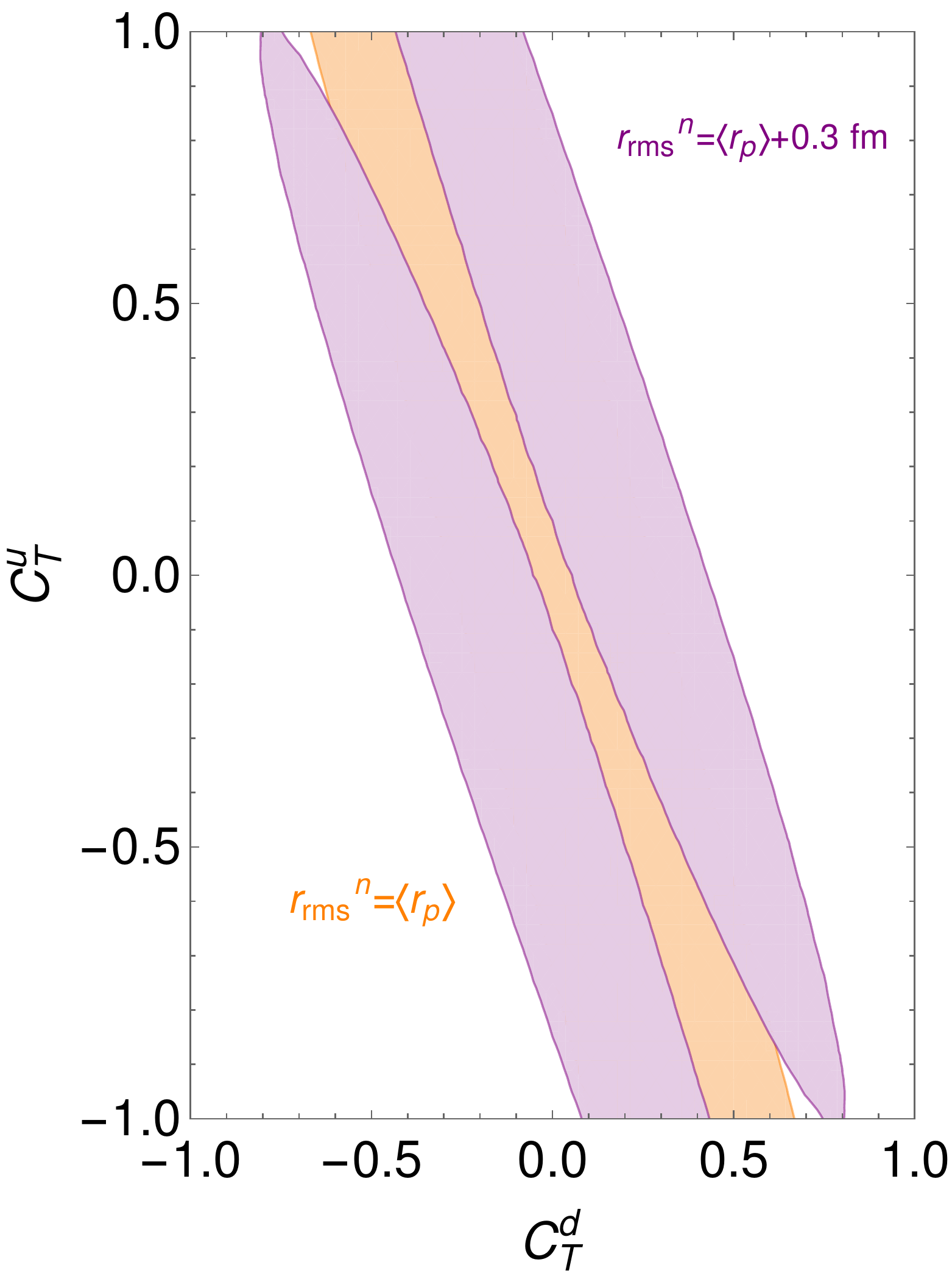}
  \caption{ 90\% C.L. allowed regions in the NGI parameter space for
    scalar interactions assuming $D_P^q=0$ ({\bf left}), for vector
    NGI assuming $D_A^q=0$ (\textbf{middle}), and for tensor NGI ({\bf
      right}). The orange bands are obtained by fixing
    $r_\text{rms}^n=\langle r_p\rangle$, while the purple regions are
    obtained for $r_\text{rms}^n=r_\text{rms}^n|_\text{max}=5.09$~fm.}
  \label{fig:ngi-results}
\end{figure*}

Finally, we turn to the discussion of the impact of the uncertainties
on the sensitivity to neutrino NGI. The results are shown in
Fig.~\ref{fig:ngi-results}. We have proceeded in the same way as that
for the NSI analysis, fixing $r_\text{rms}^n=\langle r_p\rangle$ to
generate the simulated data, and analyzing the results for two cases
with $r_\text{rms}^n=\langle r_p\rangle$ and
$r_\text{rms}^n=5.09$~fm. For scalar interactions we
assume $D_P^q=0$, while for vector interactions, $D_A^q=0$. Results in
the case $r_\text{rms}^n=\langle r_p\rangle$ are quite similar to
those found in Ref.~\cite{AristizabalSierra:2018eqm} and largely
depart from them for $r_\text{rms}^n=5.09$~fm for
reasons similar to that for NSI.  Scalar interactions are not sensitive to form factor
  uncertainties because
  fitting COHERENT data with scalar interactions leads to a
  rather poor fit, almost independently of $r_\text{rms}^n$.
$N^\text{meas}$ has been simulated assuming
$r_\text{rms}^n=\langle r_p\rangle$ and so in the presence of NGI a
better fit is found for the first sample. With
$r_\text{rms}^n=5.09$~fm there is little room for new
interactions since the mismatch between the neutron rms radii induces
substantial departure from the simulated data.
Depending on the value of $r_\text{rms}^n$ large portions of parameter
space are allowed or disfavored. 

\section{Conclusions}
\label{sec:conclusions}%
We have quantified the uncertainties on the SM CE$\nu$NS cross
section. They are driven by the neutron form factor through its
dependence on the rms radius of the neutron density distribution,
$r_\text{rms}^n$. To quantify these uncertainties we assumed that
$r_\text{rms}^n$ ranges between the rms radius of the proton charge
distribution $r_\text{rms}^p$ of the corresponding nucleus and
$r_\text{rms}^p+0.3$~fm (for heavy nuclei), so that the neutron skin is
thinner than that for $^{208}\text{Pb}$ (which has been measured by
PREX). For nuclei with $N\sim Z$, we considered $r_\text{rms}^n$ between $r_\text{rms}^p$ and $r_\text{rms}^p+0.1$~fm or $r_\text{rms}^p+0.2$~fm. Under this assumption we evaluated the size of the
uncertainties for $^{133}\text{Cs}$, $^{127}\text{I}$, germanium,
xenon and argon---choices motivated by COHERENT phases I-III and
Argo---using three form factor parameterizations: Helm, Fourier
transform of the symmeterized Fermi function and Klein-Nystrand.

We showed that form factor uncertainties: (i) are relevant for
$q\gtrsim 20\,$MeV, and so are negligible if the CE$\nu$NS process is
induced by either reactor or solar neutrinos, (ii) have percentage
uncertainties that have a strong dependence on the recoil energy (iii) are basically independent of
the parameterization used. 

We studied the impact of the uncertainties on the SM prediction for
COHERENT, diffuse supernova neutrino background, and sub-GeV
atmospheric neutrinos. For COHERENT, assuming $n_\text{POT}$/year as
in Ref.~\cite{Akimov:2017ade}, we found that the SM prediction is
subject to relative uncertainties that are never below
$1.5\%$ in germanium, $2\%$ is argon and $6\%$ in
xenon. For the combination of DSNB and atmospheric neutrinos we find that the relative
  uncertainties are at least $5\%$. These results demonstrate
that in the absence of precise measurements of $r_\text{rms}^n$, SM
predictions of the CE$\nu$NS rate involve uncertainties that challenge
the interpretation of data. This is especially true for future
measurements with small experimental systematic uncertainties.

We also quantified the impact of the neutron form factor uncertainties
on the sensitivity to new physics. We considered three scenarios:
neutrino NSI, sterile neutrinos in the 3+1 scheme, and NGI. We showed
that the variation of $r_\text{rms}^n$ has a
substantial effect on these new physics searches with the exception of scalar NGI for which we did not
  find any sensitivity.

Finally, it is worth pointing out that the uncertainties we have
derived here also apply to DM direct detection searches, provided
the WIMP-nucleus interactions are spin independent. In WIMP
scenarios with vector, scalar and tensor mediators, the
direct detection rate will involve uncertainties comparable to those
we have derived.

\section*{Acknowledgments}

We thank G.~Hagen, C.~Horowitz, J.~Piekarewicz, G.~Rich, X.~Roca-Maza, and K.~Scholberg for many useful discussions
and inputs. We also thank the organizers of the ``Magnificent
CEvNS Workshop'', where this work was initiated. This research was supported in part by the U.S. DOE
under Grant No.~de-sc0010504, by the grant ``Unraveling new physics in
the high-intensity and high-energy frontiers'', Fondecyt No 1171136,
and by the Hundred-Talent Program of Sun Yat-Sen University.

\appendix
\section{Diffuse supernova neutrino background 
  fluxes}
\label{sec:DSNB}
For the calculation of the DSNB neutrino flux we closely follow
Ref.~\cite{Horiuchi:2008jz}. Here we present the details of such a
calculation. The predicted DSNB flux is obtained by integrating the
rate of core-collapse supernova, $R_\text{SN}(z)$, multiplied by the
neutrino emission per supernova, $dN(E_\nu')/dE_\nu'$, redshifted over
cosmic time:
\begin{equation}
  \label{eq:flux-dsnb}
  \frac{d\Phi(E_\nu)}{dE_\nu}=\frac{4\pi\,c}{H_0}\,
  \int_0^{z_\text{max}}\,R_\text{SN}(z)\,
  \frac{dN_{\nu_i}(E_\nu')}{dE_\nu'}\,
  \frac{dz}{\sqrt{\Omega_m(1+z)^3 + \Omega_\Lambda}}\ .
\end{equation}
Here the redshifted neutrino energy is given by $E_\nu'=(1+z)E_\nu$,
and $z_\text{max}$ is determined by gravitational collapse, assumed to
start at $z=5$ \cite{Ando:2004hc}. The cosmological parameters have been
fixed to: $\Omega_m=0.3$, $\Omega_\Lambda=0.7$ and
$H_0=70\,\text{km}\,\text{s}^{-1}\,\text{Mpc}^{-1}$. The rate for
core-collapse supernova in units of Mpc$^{-3}$year$^{-1}$ is
determined by the star formation rate $\dot\rho_*(z)$ and the initial
mass function $\Psi(M)$ as
\begin{equation}
  \label{eq:rate-for-core-collapse-SN}
  R_\text{SN}(z)=\dot\rho_*(z)\,
  \frac{\int_{8\text{M}_\odot}^{150\text{M}_\odot}
    \Psi(M)dM}{\int_{0.1\text{M}_\odot}^{100\text{M}_\odot}M\Psi(M)dM}\ .
\end{equation}
The integral in the numerator gives the number of stars that produce
core collapse supernova, while the integral in the denominator gives the
total mass in stars. The initial mass function $\Psi(M)=dN/dM$
determines the number of stars with masses in the range $M$ and $M+dM$
and reads $\Psi(M)=M^{-\xi}$, with the value of $\xi$ defining a
particular initial mass function and therefore the value of the
integrals in (\ref{eq:rate-for-core-collapse-SN}). For our calculation
we have used a Baldry-Glazebrook initial mass function
\cite{Baldry:2003xi} for which the integral has a value of
$0.0132/\text{M}_\odot$. The star formation rate is given by the
fitted function
\begin{equation}
  \label{eq:fitted-SFH}
  \dot\rho_*(z)=\dot\rho_0
  \left[
    (1+z)^{\alpha\eta}
    +
    \left(\frac{1+z}{B}\right)^{\beta\eta}
    +
    \left(\frac{1+z}{C}\right)^{\gamma\eta}
  \right]^{1/\eta}\ ,
\end{equation}
where $\eta=10$ and the constants $B$ and $C$ are
\begin{equation}
  \label{eq:fitted-SFH}
  B=(1+z_1)^{1-\alpha/\beta}\ ,\qquad
  C= (1+z_1)^{(\beta-\alpha)/\gamma}(1+z_2)^{1-\beta/\gamma}\,.
\end{equation}
 We fix $\alpha=3.4$, $\beta=-0.3$, $\gamma=-3.5$, $z_1=1$,
$z_2=4$ and
$\dot\rho_0=0.0178\,\text{M}_\odot\text{year}^{-1}\text{Mpc}^{-3}$,
which correspond to the fiducial analytic fit given in Ref.~\cite{Horiuchi:2008jz}.

In core-collapse supernova, neutrinos of all flavors are emitted and
each flavor carries about the same fraction of the total energy,
$E_\nu\simeq 3\times 10^{53}$~erg. Their spectra are approximately
thermal with temperatures obeying $T_{\bar\nu_e}<T_{\nu_e}<T_{\nu_x}$
($\nu_x=\nu_\mu,\nu_\tau,\bar\nu_\mu,\bar\nu_\tau$). We have taken a
Fermi-Dirac distribution with zero chemical potential for all flavors,
\begin{equation}
  \label{eq:Fermi-Dirac}
  \frac{dN_{\nu_i}(E'_\nu)}{dE_\nu}=\frac{E_\nu^\text{tot}}{6}
  \frac{120}{7\pi^4}\frac{E_{\nu_i}^{'2}}{T_{\nu_i}}
  \frac{1}{e^{E_{\nu_i}'/T_{\nu_i}} + 1}\ ,
\end{equation}
and temperatures according to: $T_{\bar\nu_e}=3$ MeV, $T_{\nu_e}=5$
MeV and $T_{\nu_x}=8$ MeV.

\section{NGI cross section parameters}
\label{sec:parameters-def}
The parameters in Eq.~(\ref{eq:xsecs-NGI}) are closely related to those in
Ref.~\cite{AristizabalSierra:2018eqm}, but involve a $q^2$ dependence
related to the proton and neutron form factors. For the $\xi_X$
($X=S,V,T$) couplings we have
\begin{align}
  \label{eq:xi-param}
  \xi_S^2(q^2)&=C_S^2(q^2)+D_P^2(q^2)\ ,
  \qquad
  \xi_T^2(q^2)=2 C_T^2(q^2)\ ,
  \nonumber\\
  \xi_V^2(q^2)&=\left[C_V^2(q^2)+2g_V(q^2)\right]^2+D_A^2(q^2)\ ,  
\end{align}
with $g_V=g_V^nF_n(q^2)+g_V^pF_p(q^2)$.  In the scalar case, the
parameters that define $\xi_S$ read
\begin{equation}
  \label{eq:scalar-CS}
  C_S(q^2)=\sum_{q=u,d}C_s^{(q)}\left[N\frac{m_n}{m_q}f_{T_q}^nF_n(q^2)
  + Z\frac{m_p}{m_q}f_{T_q}^pF_p(q^2)\right]\ ,
\end{equation}
and the same definition applies for $D_P(q^2)$ by trading
$C_S^{(q)}\to D_P^{(q)}$. The parameters $f_{T_q}$ are derived in
chiral perturbation theory from measurements of the $\pi$-nucleon
sigma term \cite{Cheng:1988im}. For our calculations we use the values
\begin{alignat}{2}
  \label{eq:ftq-values}
  f_{T_u}^p&=0.019\ ,\qquad&f_{T_d}^p=0.041\ ,
  \nonumber\\
  f_{T_u}^n&=0.023\ ,\qquad&f_{T_d}^n=0.034\ .
\end{alignat}
For the vector coupling $\xi_V$ we have
\begin{equation}
  \label{eq:vector-CV}
  C_V(q^2)=N(C_V^u+2C_V^d)F_n(q^2) + Z(2C_V^u+C_V^d)F_p(q^2)\ .
\end{equation}
The expression for $D_A(q^2)$ can be obtained from
(\ref{eq:vector-CV}) by trading $C_V^{q}\to D_A^{q}$ with $q=u,\,d$.  Finally, in
the tensor case,
\begin{equation}
  \label{eq:tensor-CT}
  C_T(q^2)=N(\delta_u^nC_T^u+\delta_d^nC_T^d)F_n(q^2) 
  + Z(\delta_u^pC_T^u+\delta_d^pC_T^d)F_p(q^2)\ .
\end{equation}
For $\delta_q^n$ and $\delta_q^p$ we use values obtained from
azimuthal asymmetries in semi-inclusive deep-inelastic scattering and
$e^+e^-\to h_1h_2X$~\cite{Anselmino:2008jk}; more up-to-date
values can be found in
\cite{Courtoy:2015haa,Goldstein:2014aja,Radici:2015mwa}. For our
calculation we use
\begin{alignat}{2}
  \label{eq:delta-values}
  \delta_u^p&=0.54\ ,\qquad&\delta_d^p=-0.23\ ,
  \nonumber\\
  \delta_u^n&=-0.23\ ,\qquad&\delta_d^n=0.54\ .
\end{alignat}
\bibliography{references}
\end{document}